\documentclass[12pt,letterpaper]{article}
\usepackage{jheppub,amsmath}
\usepackage{natbib}
\usepackage{graphicx}
\usepackage{subfig}
\usepackage{amssymb}
\usepackage{tensor}
\usepackage{physics}

\makeatletter
\def\@fpheader{\relax}
\makeatother
\subheader{}
\title{Quantum Information Bound on the Energy}

\author[a,b]{Raphael Bousso,}
\affiliation[a]{Center for Theoretical Physics and Department of Physics,\\
 University of California, Berkeley, CA 94720, U.S.A.}
\affiliation[b]{Lawrence Berkeley National Laboratory, Berkeley, CA 94720,
  U.S.A.}

\author[a,b]{Arvin Shahbazi-Moghaddam,}

\author[c]{and Marija Toma\v{s}evi\'{c}}
\affiliation[c]{Departament de F\'isica Qu\`antica i Astrof\'isica and Institut de Ci\`encies del Cosmos (ICCUB),\\
Universitat de Barcelona, Mart\'i i Franqu\`es 1, E-08028 Barcelona, Spain}

\abstract{According to the classical Penrose inequality, the mass at spatial infinity is bounded from below by a function of the area of certain trapped surfaces. We exhibit quantum field theory states that violate this relation at the semiclassical level. We formulate a Quantum Penrose Inequality, by replacing the area with the generalized entropy of the lightsheet of an appropriate quantum trapped surface. We perform a number of nontrivial tests of our proposal, and we consider and rule out alternative formulations. We also discuss the relation to weak cosmic censorhip.}

\begin{document}
\maketitle

\section{Introduction}
\label{sec-intro}

Semiclassical General Relativity allows for quantum matter while keeping the gravitational field classical, by coupling the metric to the expectation value of the stress tensor:
\begin{equation}
  G_{ab} = 8\pi G \langle T_{ab}\rangle~.
\end{equation}
Since $\langle T_{ab}\rangle$ receives quantum contributions proportional to $\hbar$, this approximation can be organized as a perturbative expansion in $G\hbar$ and solved iteratively. This approach has proven to be quite useful, leading to the discovery of black hole thermodynamics and the associated information paradox.

Numerous theorems in General Relativity rely on the Null Energy Condition (NEC), which states that 
\begin{equation}
  T_{ab} k^a k^b\geq 0
  \label{eq-nec}
\end{equation}
at every point in the spacetime, where $k^a$ is any null vector. The NEC underlies the area theorems for event horizons~\cite{Haw71} and for future holographic screens~\cite{BouEng15a,BouEng15b}, the focussing theorem~\cite{Wald}, and Penrose's singularity theorem~\cite{Pen64}. In other theorems, the stress tensor is assumed to obey even stronger conditions, which are nevertheless satisfied by known classical matter and radiation.

However, in relativistic quantum field theories (QFTs) such as the Standard Model, there are states in which $\langle T_{ab} \rangle$ violates the NEC in some regions of spacetime. Hence, none of the classical theorems mentioned above apply at the semiclassical level. The evaporation of a black hole, for example, is accompanied by violations of all of the above theorems. This is possible because the NEC is violated in the vicinity of the horizon. 

Remarkably, there is considerable evidence that all of the above theorems admit a conjectural semiclassical extension. The key step to obtaining a viable proposal is to replace the area of surfaces with their generalized entropy. Thus the area theorem becomes
the Generalized Second Law (GSL) for event horizons~\cite{Bek72,Bek73,Bek74} and for Q-screens~\cite{BouEng15c}. The focussing theorem becomes the Quantum Focussing Conjecture (QFC) ~\cite{BouFis15a}; and Penrose's singularity theorem becomes Wall's Quantum Singularity Theorem~\cite{Wal13}.

Though these are conjectural statements about the semiclassical limit of quantum gravity, they can have interesting nongravitational limits. Some of these limit statements were already known, but others came as completely new and nontrivial results in QFT. The main example is the Quantum Null Energy Condition~\cite{BouFis15a}, which has since been rigorously proven within QFT, using a variety of methods~\cite{BouFis15b,KoeLei15,BalFau17}. Thus, the study of semiclassical gravity has had considerable impact in a seemingly unrelated arena.

The present work is inspired by these developments. We will study an important conjecture in classical General Relativity, the Penrose inequality~\cite{PenNS}.  This a relation between the area of certain marginally trapped surfaces $\mu$ in the spacetime and the total mass defined at spatial infinity~\cite{ADM}:
\begin{equation}
  m\geq \sqrt{\frac{A[\mu]}{16\pi G^2}}~.
  \label{eq-piintro}
\end{equation}
The conjecture can be thought of as a generalization of the positive mass theorem~\cite{SchYau81}. For either statement, it is clearly essential that matter with negative energy be excluded. This can be implemented by assuming the dominant energy condition (DEC), that for any timelike future-directed vector $t^a$, $-T^a_{~b}t^b$ is timelike and future-directed.

The Penrose inequality has not been proven and thus is not a theorem. But no counterexample to the conjecture is known. We will review the classical Penrose inequality in Sec.~\ref{sec-CPI}, where we provide both the reasoning motivating it, and a more careful formulation.

Since quantum matter can violate the NEC it can also violate the DEC, threatening the validity of the Penrose inequality. It is not immediately obvious that Eq.~(\ref{eq-piintro}) fails, since the stress tensor in QFT cannot be dialed arbitrarily.

In fact, we find in Sec.~\ref{sec-boulware} that Eq.~(\ref{eq-piintro}) continues to be satisfied in a simple example of black hole formation and evaporation. However, we then provide an explicit counterexample to the classical Penrose inequality, by exploiting the thermal nature of the vacuum state near the horizon. When the thermal state is depleted, the vicinity of the horizon can contribute significant negative energy. This cancels an order one fraction of the black hole's mass, leading to a substantial violation of Eq.~(\ref{eq-piintro}).

We are thus motivated, in Sec.~\ref{sec-QPI}, to propose a quantum-corrected version of the Penrose inequality. We introduce the relevant concepts of generalized entropy, quantum expansion, and quantum (marginally) trapped surfaces. We draw some lessons from the failure of the classical Penrose inequality in the semiclassical setting, and we formulate a Quantum Penrose Inequality (QPI).

In Sec.~\ref{sec-evidence}, we provide evidence for our proposal. We consider several interesting examples that could challenge the QPI, and we show that our proposal survives these tests. In Sec.~\ref{sec-alternatives}, we discuss a number of alternative formulations of the QPI. We show why they are either excluded or not ideal. In Sec.~\ref{sec-ads}, we discuss the formulation of a QPI in asymptotically Anti-de Sitter spacetimes. This helps us identify subtleties that also affect the original QPI.

In Sec.~\ref{sec-limits}, we discuss the classical and the nongravitational limits of the QPI.

Penrose's motivation in proposing Eq.~(\ref{eq-piintro}) was as a test of the weak Cosmic Censorship Conjecture (CCC). In Sec.~\ref{sec-ccc}, we review this connection and the status of the CCC. We speculate that the Quantum Penrose Inequality could inform the formulation of a ``quantum'' CCC that accommodates the known, physically sensible violations of the classical CCC. 

In Appendix~\ref{sec-geom}, we compute the expansion of outgoing null rays and the positions of classical and quantum marginally trapped surfaces for an evaporating Schwarzschild black hole. In Appendix~\ref{sec-qscreens}, we present a perturbative construction of Q-screens~\cite{BouEng15c}, which plays a role in our discussion of the Quantum Penrose Inequality in Anti-de Sitter spacetimes.

A brief summary of the main results of our investigation has appeared elsewhere~\cite{BouSha19a}.

\section{Classical Penrose Inequality}
\label{sec-CPI}

In this section we describe the (classical) Penrose inequality; see Ref.~\cite{Mar09} for a broader review.

\subsection{Formulation}
\label{sec-cnd}

We formulate the classical Penrose inequality as follows:

Let $m$ be the total mass of an asymptotically flat spacetime. Let $\mu$ be a trapped surface that has minimal area among all surfaces that enclose it, on some Cauchy surface that contains $\mu$. Then 
\begin{equation}
  m\geq \sqrt{\frac{A[\mu]}{16\pi G^2}}~.
  \label{eq-pi}
\end{equation}
Next, we provide detailed definitions and explanations of the terms appearing in this formulation.

Let $(M,g_{ab})$ be a connected Lorentzian spacetime with metric. Let $\mu$ be a codimension $1+1$ compact spacelike submanifold (a ``surface'').~\footnote{In the remainder of this paper we will specialize to 3+1 dimensional spacetime, so that $\mu$ will be a 2-dimensional surface. Generalization to higher dimensions is trivial.} Let $\theta_\pm$ be the expansion of the future-directed light-rays emanating orthogonally from $\mu$ to either side. If $\theta_+\leq 0$ and $\theta_-\leq 0$ then $\mu$ is called {\em trapped}. If $\theta_+=0$ and $\theta_-\leq 0$ then $\mu$ is {\em marginally trapped}.

Now let $(M,g_{ab})$ be in addition asymptotically flat.
Note that we do not require $\mu$ to be connected; for example in a spacetime where multiple black holes are forming, $\mu$ could be the union of connected marginally trapped surfaces inside some or all of them.

Suppose that the surface $\mu$ has an ``outer wedge'' $O_W$ that contains a single asymptotic region. By this we mean that $\mu$ forms the only boundary of any Cauchy surface of a globally hyperbolic region of space $O_W$ that (in the ``unphysical spacetime'' or Penrose diagram) contains a single copy of spatial infinity, $i_0$. This will be the case for trapped surfaces in a spacetime with a single asymptotic region. In the case of ``two-sided'' black hole solutions, it will hold if $\mu$ is homologous\footnote{Two cycles (closed submanifolds which are not boundaries of any other submanifolds) are said to be homologous, or equivalently, belong to the same homology class, if they can be continuously deformed into each other.} to a horizon (with either choice of side), but not if $\mu$ is contractible. We will be interested in bounding the mass at spatial infinity~\cite{ADM} from below.

Finally, we assume that there exists a Cauchy surface $\Sigma$ of $O_W$ on which $\mu$ is the minimal area surface homologous to large spheres near $i_0$ (or in the AdS case, homologous to the boundary sphere)~\cite{EngHor19}. The purpose of this set of assumptions will become clear as we turn to presenting a heuristic argument that the Penrose Inequality should hold for $\mu$.

\subsection{Heuristic Argument}

The Penrose Inequality was originally intended as a test of cosmic censorship, which guarantees that an asymptotically flat spacetime with regular initial conditions will be strongly asymptotically predictable~\cite{Wald}. If this latter property holds, then a compelling argument can be given that the Penrose inequality must hold; thus, any regular initial data set that violates the Penrose inequality would likely exclude cosmic censorship. We now present the argument.

Roughly speaking, strong asymptotic predictability establishes the existence of $\tilde V$, a globally hyperbolic open subset of $M$ that contains any black hole horizons and their exterior, $\tilde V\supset \bar{J^-({\cal I}^+)}$. (See Ref.~\cite{Wald} for more details.) The {\em black hole region} is $B\equiv M-J^-({\cal I}^+)$. The black hole {\em event horizon} is its boundary $\dot B$.

Suppose that
\begin{equation}
  R_{ab}k^ak^b\geq 0~,
  \label{eq-ncc}
\end{equation}
as would be the case if Einstein's equations hold with matter satisfying the Null Energy Condition. Then any trapped or marginally trapped surface $\mu$ must lie in the black hole region:
\begin{equation}
  \mu\subset B~.
\end{equation}
For a proof, see Propositions 12.2.2 in Ref.~\cite{Wald}.  The key technical assumption is that $M$ be strongly asymptotically predictable.\footnote{The same property, $\nu\subset B$, follows from Proposition 12.2.3 in Ref.~{Wald} for another class of surfaces called {\em outer trapped}. These would form an alternate starting point from which the classical and quantum Penrose conjectures could be developed along the same lines as we do here for trapped surfaces.}

Let $H=\dot B\cup \Sigma$ be the slice of the black hole event horizon (possibly with multiple disconnected components), on the Cauchy surface $\Sigma$ of $O_W$. Since $\mu$ has minimal area on $\Sigma$, it follows that the horizon must be at least as large:\footnote{Instead of assuming that $\mu$ has minimal area on {\em some} Cauchy slice of $O_W$, an alternative way of handling this issue is to replace $A[\mu]$ with the minimal area of all surfaces enclosing $\mu$ on a {\em given} initial Cauchy slice~\cite{Mar09}. Verifying this assumption does not require knowledge of more than the initial slice.}
\begin{equation}
  A[H]\geq A[\mu]~.
  \label{eq-hmn}
\end{equation}

The Null Curvature Condition, Eq.~(\ref{eq-ncc}), and strong asymptotic predictability imply that the area of the event horizon cannot decrease with time~\cite{Haw71}. Let $H'=\dot B\cup\Sigma'$, where $\Sigma'$ is a Cauchy surface to the future of $\Sigma$. Then
\begin{equation}
  A[H']\geq A[H]~.
  \label{eq-hph}
\end{equation}

Physically, it is reasonable to assume that regular initial data will eventually settle down to a Kerr black hole. (In four dimensions, this follows from the assumption of late-time stationarity, by the Israel-Hawking-Carter theorems~\cite{HawEll}.) Letting $H'$ be a slice of the horizon at this late time, the formula for the area of a Kerr black hole implies that
\begin{equation}
  16 \pi G^2 m_{\rm Kerr}^2\geq A[H']~.
\label{eq-hpm}
\end{equation}
The spacetime will not be exactly Kerr, however. One expects that massive fields will have fallen into the black hole, but there may be massless fields that propagate to future null infinity. Because this radiation becomes dilute and well separated from the black hole, gravitational binding energy will be negligible. Hence the ADM mass, $m$, will be given by the sum
\begin{equation}
  m = m_{\rm Kerr} + m_{\rm rad}\geq m_{\rm Kerr}~.
\label{eq-mgm}
\end{equation}
Combining the previous four inequalities, we obtain the Penrose conjecture, Eq.~(\ref{eq-pi}).

We would like to add a second, somewhat independent heuristic argument for Eq.~(\ref{eq-pi}). A \textit{future holographic screen} is a hypersurface foliated by marginally trapped surfaces called {\em leaves}~\cite{CEB1,CEB2,BouEng15a}. Assuming the Null Energy Condition, the area of the leaves increases monotonically along this foliation~\cite{BouEng15a,BouEng15b}. In the spherically symmetric case, the screen eventually asymptotes to the event horizon (from the interior), so its final area will be equal to the late time event horizon area. Thus the screen area theorem implies the Penrose inequality in this case. More generally, given a marginally trapped surface $\mu$, a future holographic screen can be constructed at least in a neighborhood. The Penrose inequality would follow from the stronger assumption that there exists a future holographic screen that interpolates from $\mu$ to the late-time event horizon, as in the spherical case.

\section{Violation by Quantum Effects}
\label{sec-boulware}

In this section, we will show that there is a need for a quantum generalization of the classical Penrose inequality (CPI). We will construct an explicit counterexample that is based on a Boulware-like state outside a Schwarzschild black hole. It violates the CPI by a substantial, classical amount.

This will be a counterexample to the CPI in the same sense as black hole evaporation is a counterexample to Hawking's area theorem: we identify a physically allowed state in which a key assumption of the classical statement, the Null Energy Condition, does not hold, and we verify that the conclusion fails as well.

\begin{figure}%
     \centering
     \includegraphics[width=1\textwidth]{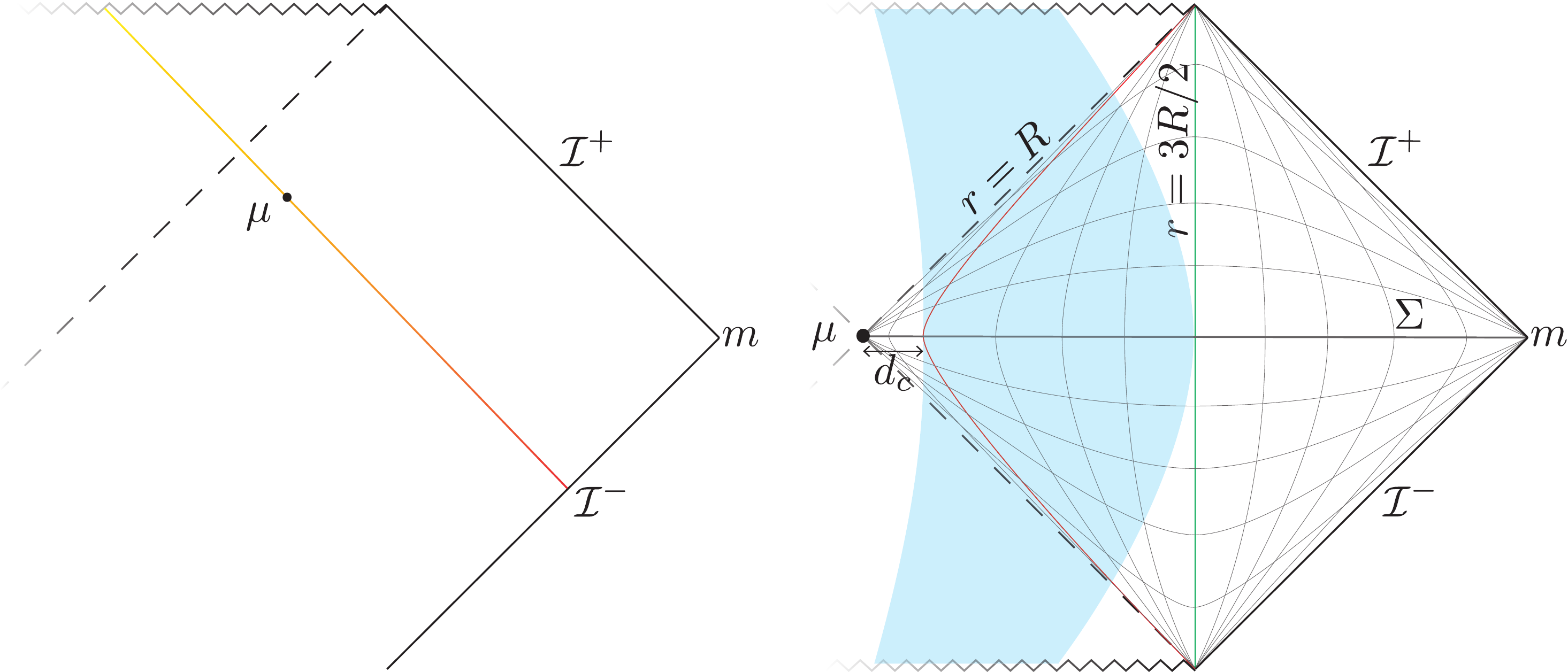}
     \caption{Left: A null shell collapsing in asymptotically flat spacetime. The classically marginally trapped surface $\mu$ is slightly outside of the event horizon due to the evaporation. It is not clear that this example violates the CPI. Right: initial data that violates the classical Penrose inequality. Here $\mu$ is the bifurcation surface of the Schwarzschild (Kruskal) solution. Inside a proper distance $d_c$, the state is the Hartle-Hawking vacuum. Outside of $d_c$,it becomes the Boulware vacuum, which has negative energy in the near-horizon zone (blue strip). This lowers the mass at infinity by an $O(1)$ fraction compared to a classical black hole.}%
     \label{fig-flatBoulware}%
\end{figure}

However, before we turn to our counterexample, it is worth noting that no obvious violation of the CPI arises in the ``normal'' formation and evaporation of a black hole in the Unruh state. This is interesting, because in this setting the Null Energy Condition is already violated, and other theorems like the area theorem or the focussing theorem do fail. In order to have full control and exclude transient effects, let us consider the collapse of a null shell of mass $m$; see Fig.~\ref{fig-flatBoulware}. Then by causality, there are no corrections to the classical solution on the shell and to its past, where the spacetime is a portion of Minkowski space. In particular, the marginally trapped surface on the shell will have the same area as in the classical case, and the CPI will be saturated. (The fact that the event horizon is inside of this surface is irrelevant.) At later times, we expect the apparent horizon area to decrease. Since the mass at infinity does not change during evaporation, the CPI will remain satisfied.

We do not claim that the CPI will hold for all black holes formed from collapse; and even in the above example, its validity may rely on idealizations, such as treating the collapsing null shell as infinitely thin and stable. But we would like to exhibit a situation where the CPI is definitely violated; in order to do this, we will consider a somewhat more artificial (but certainly valid) quantum state.

To demonstrate a violation of the classical PI by quantum effects, we now consider a Boulware-like state~\cite{Boulware} of a massless scalar field, on one side of a maximally extended Schwarzschild black hole, at the time-symmetric slice; see Fig.~\ref{fig-flatBoulware}. The Boulware vacuum is analogous to the Rindler vacuum. It corresponds to vanishing occupation number of the modes with support strictly outside the event horizon. This will contribute some negative energy outside of the black hole, in the near-horizon region $R<r<3R/2$. Far from the black hole, the stress tensor vanishes in the Boulware vacuum.

Note that the classical Penrose inequality, applied to the bifurcation surface, is classically saturated. (That is, it is saturated if the stress tensor vanishes everywhere outside the black hole.) Thus, any net negative energy in the exterior will lead to a violation of Eq.~(\ref{eq-pi}). 

The local stress tensor diverges in the Boulware vacuum as the horizon is approached~\cite{Boulware,Candelas}. We regulate this divergence by building wavepackets with support strictly outside of a sphere $H_c$ at proper distance $d_c>0$ from the horizon (in this case, from the bifurcation surface). For full control of the semiclassical expansion, we choose
\begin{equation}
  l_P\ll d_c \ll R~.
  \label{eq-regime}
\end{equation}
Roughly speaking, this yields a Hartle-Hawking-like state (vanishing stress tensor) inside of $H_c$, and a Boulware-like state outside of $H_c$.

Integration of the QFT stress tensor computed in Ref.~\cite{Candelas}, outside the regulator sphere $H_c$, yields a QFT contribution to the energy at infinity of order $-(l_P/d_c)^2 M$, where $M=R/2G$ is the mass of the black hole~\cite{BouSha19a}. Here we will go further; instead of naively gluing to QFT states across a surface (which is does not generally yield an allowed QFT state), we consider junction effects at $H_c$. Positivity of the energy for infalling observers requires some positive energy near $H_c$, which we wish to estimate and show to be negligible.

For this purpose it will be useful to analyze the problem mode by mode. This will allow us to distinguish between two cutoffs that we can freely choose: the angular momentum of the included QFT modes, and $d_c$. Establishing a small hierarchy between these cutoffs will give us a control parameter $1/n_{\rm node}\ll 1$, by which the positive energy at $H_c$ is suppressed at infinity, relative to the negative contribution.

We will focus on the most relevant modes in the near-horizon zone, which have occupation number of order one in the thermal ensemble corresponding to the Hartle-Hawking state. These modes have the property that any wavepacket constructed from them has characteristic wavelength comparable to its distance from the horizon. Moreover, increasing the occupation number of the mode by 1 increases the energy at infinity by $\hbar/R$.
\begin{figure}%
     \centering
     \includegraphics[width=.6\textwidth]{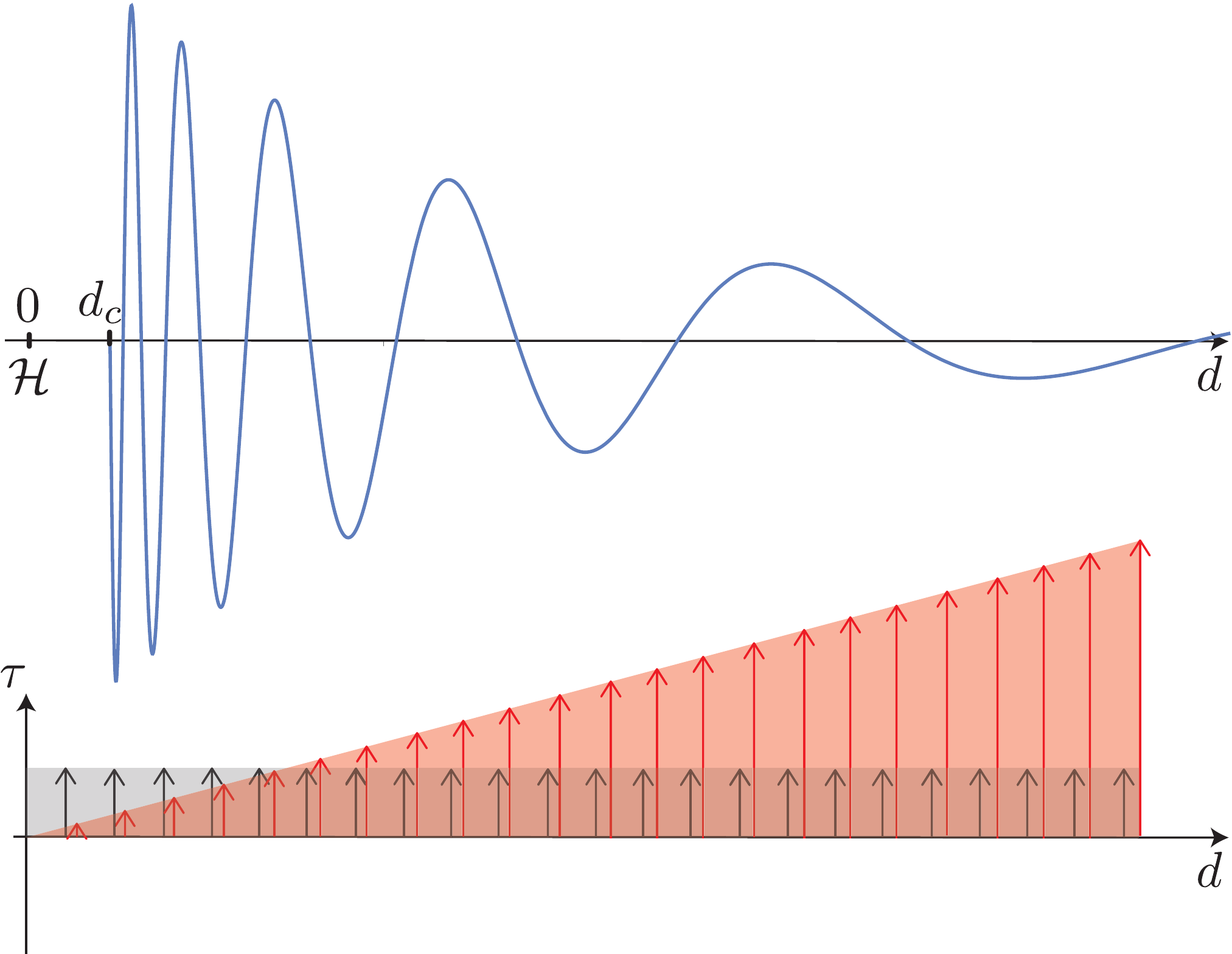}
     \caption{A typical wavepacket mode in the thermal atmosphere of the black hole, regulated to have support outside a sphere a proper distance $d_c$ outside of the horizon. The classical Penrose inequality is violated in a Boulware-like state in which such modes have zero occupation number and negative energy. In a local inertial frame (black Killing vector field, $\partial_\tau$, where $\tau$ is proper time), a large fraction of their energy is concentrated near the cutoff $d_c$. The total energy must appear positive in this frame; this can be satisfied by adding a comparable amount of positive energy inside of $d_c$. To an asymptotic observer (red Killing vector field, $\partial_t$), the negative energy is spread evenly over the mode, due to the greater redshift near the horizon. Thus the positive energy beyond the cutoff has a negligible effect on the ADM mass. }%
     \label{fig-Rindlermode}%
\end{figure}

This set of modes includes $s$-waves as well as modes with nonzero angular momentum. Here we will use $\ell=0,1,\ldots$ for the angular momentum quantum number. The number of modes in the thermal atmosphere can be estimated from the number of nodes in a strictly outgoing Rindler mode in an interval beginning at proper distance $d_c$ from the horizon and ending at a distance $R$ (for the spherical modes, which we approximate as propagating freely) or $R/(\ell+1)$ (for the modes with angular momentum, which we approximate as being reflected by an angular momentum barrier). See Fig.~\ref{fig-Rindlermode}. Hence there are
\begin{equation}
 n_\ell= (2\ell+1) \log\left(\frac{R/(\ell+1)}{d_c}\right)
\end{equation}
linearly independent modes with angular momentum $\ell$. 

In the Hartle-Hawking state, these modes are all thermally excited with $O(1)$ occupation numbers; this corresponds to vanishing stress tensor near the horizon. In the Boulware-like state, the modes are unoccupied. This corresponds to a negative stress tensor; it contributes an energy at infinity of order $-\hbar/R$, per mode. We choose a cutoff $\ell_{\rm max}$ on the angular momentum such that the angular momentum barrier is somewhat outside the short distance cutoff $d_c$:
\begin{equation}
  \log \log\left( \frac{R/(\ell_{\rm max}+1)}{d_c}\right) \sim O(1)~,
  \label{eq-cutoffs}
\end{equation}
where the second $\log$ enforces a small hierarchy whose purpose will become clear below. From the previous two equations, the total number of unoccupied modes is
\begin{equation}
  n_{\rm total} \equiv \sum_{\ell=0}^{\ell_{\rm max}} n_\ell \sim \frac{R^2}{d_c^2}~.
\end{equation}
Thus the total energy at infinity of the quantum field will be
\begin{equation}
  E_{\rm neg}\sim  - \frac{\hbar}{R} n_{\rm total} \sim - \alpha M~,
  \label{eq-eneg}
\end{equation}
where
\begin{equation}
  \alpha = \frac{l_P^2}{d_c^2}~.
  \label{eq-alpha}
\end{equation}

The presence of a substantial amount of negative energy outside the black hole may seem suspect. However, we note that our construction cannot achieve vanishing or negative total ADM mass. Since the black hole contributes $M$, the total mass is $(1-\alpha)M$. Making this negative would require taking $d_c\lesssim l_P$, in conflict with Eq.~(\ref{eq-regime}), and so would take us outside of the semi-classical expansion. Moreover, our result is consistent with positive total matter energy in an appropriate neighborhood of the horizon. This is important since the spacetime can be treated as approximately flat on a distance scale $d_c\ll d_{\rm flat}\ll R$.

To see this, we note that the wavepackets we study have approximately constant Killing energy per cycle, where a cycle denotes the portion of a wavepacket between two nodes. See Fig.~\ref{fig-Rindlermode}. The local proper wavelength of a given mode grows as the distance from the horizon, but this is precisely cancelled by the decreasing redshift. Thus from the viewpoint of infinity, each cycle of each mode contributes an ADM energy (per occupation number) of $\hbar/(R n_{\rm node})$, where
\begin{equation}
 n_{\rm node}(\ell)\sim  \log\left(\frac{R/(\ell+1)}{d_c}\right)
\end{equation}
is the number of nodes or cycles in the wavepacket. 

In a local inertial frame, on the other hand, there is no redshift effect. Yet, the proper wavelength grows exponentially away from the horizon, roughly doubling with every cycle. Thus an $O(1)$ fraction of the local energy of a mode is contained in the first phase cycle. In the Boulware-like state, this is the negative energy that must be cancelled. To have positive energy in the local frame, it suffices to have compensating positive energy just for this first cycle. The positive energy can be localized, for example, just below $d_c$.

This positive energy will partially cancel the negative ADM energy of the quantum state, Eq.~(\ref{eq-eneg}). But because all cycles of the wavepacket contribute equally to the Killing energy, the correction is parametrically small, of order $|E_{\rm neg}|/n_{\rm node}\ll |E_{\rm neg}|$. In practice, $n_{\rm node}$ of order a few suffices, so we will not update Eq.~(\ref{eq-alpha}). The purpose of the second log in Eq.~(\ref{eq-cutoffs}) was to chose the angular momentum cutoff $\ell_{\rm max}$ so as to achieve $n_{\rm node}\sim$ a few, for all modes involved in the construction. 

Finally, we note that the location and area of the marginally trapped surface do not receive large enough corrections to rescue the classical Penrose inequality. The bifurcation surface remains marginally trapped when we pass from the classical treatment to the Hartle-Hawking state, since the stress tensor vanishes there. Our construction keeps the Hartle-Hawking state near the bifurcation surface, up to corrections that can be suppressed arbitrarily by dialing $n_{\rm node}\gg 1$.

To summarize, one can reduce the mass at infinity from $M$ (in the Unruh state) to $(1-\alpha) M$ in the Boulware-like state. Since we require that $l_P\ll d_c$ for control, this correction is parametrically small, $\alpha\ll 1$. But since the Penrose inequality is saturated classically for a Schwarzschild black hole, our example violates it.

Moreover, the violation is substantial in the sense that it is not $O(\hbar)$ but $O(1)$. The contribution from each mode is $O(\hbar)$; but the number of available modes in the thermal atmosphere, at fixed control parameter $l_P/d_c$, is $n_{\rm total}\sim O(\hbar^{-1})$. Thus, the negative energy of the quantum fields can cancel off an $O(1)$ fraction of the black hole's classical mass.

\section{Quantum Penrose Inequality}
\label{sec-QPI}

In this section, we will formulate the Quantum Penrose Inequality (QPI). In Sec.~\ref{sec-qnd}, we review various concepts necessary for the quantum generalization of classical statements involving area and null expansion. In Sec.~\ref{sec-motivation}, we draw some conclusions from the failure of the classical Penrose inequality. In Sec.~\ref{sec-formulation}, we formulate our proposal for the QPI. 

\subsection{Generalized Entropy and Quantum Expansion}
\label{sec-qnd}

We begin by introducing the notion of generalized entropy and its main properties. We then use the generalized entropy to define certain quantum generalizations of various geometric quantities, necessary for formulating the Quantum Penrose Inequality; see ~\cite{BouFis15a} for more details.

The \textit{generalized entropy} $S_{\text{gen}}$, was first introduced by Bekenstein \cite{Bek72,Bek73} as the total entropy of a system consisting of a black hole and its exterior on a given time slice. The definition can be extended to apply not only to the horizon of a black hole, but to any Cauchy-splitting surface $\sigma$:
\begin{equation}
  S_{\text{gen}} \equiv \frac{A[\sigma]}{4G\hbar} + S_{\text{out}} + \dots~,
  \label{eq-gedef}
\end{equation}
where $A[\sigma]$ is the area of $\sigma$, and
\begin{equation}
  S_{\text{out}}= - {\rm Tr} \rho_{\rm out} \log  \rho_{\rm out} 
\end{equation}
is the von Neumann entropy of the state of the quantum fields, restricted to one side of $\sigma$:
\begin{equation}
\rho_{\rm out} = {\rm Tr}_{\overline{\rm out}}\, \rho~.
\end{equation}
Here, the state $\rho$ is the global quantum state, and the trace is over the complement region, which we define as $\overline{\rm out}$.

The von Neumann entropy $S_{\rm out}$ quantifies the amount of entanglement in the vacuum across $\sigma$, and as such, has divergences coming from short-distance entanglement. The leading divergence is given by $A/\epsilon^2$, where $\epsilon$ is a short-distance cutoff. However, we can think of the geometric term in Eq.~(\ref{eq-gedef}) as a counterterm. The dots indicate the presence of subleading divergences in $ S_{\rm out}$ which come with their own geometric counterterms. It is expected that the divergences coming from the renormalization of $G$ and from short-distance entanglement will cancel out~\cite{BouFis15a}, so as to keep $S_{\text{gen}}$ a finite and well-defined quantity.

One can interpret $S_{\text{gen}}$ in two distinct ways.  Following the original motivation, one can view the area-term as a (large) ``correction'' to the entropy of quantum fields. Alternatively, we can define a quantum-corrected area of the surface $\sigma$:
\begin{equation}
  A_Q[\sigma] \equiv A[\sigma]+4G\hbar S_{\text{out}} + \dots~,
  \label{eq-quanarea}
\end{equation}
in a semiclassical expansion in $G\hbar$. Hence, one can use the notion of generalized entropy to incorporate quantum effects into certain geometrical objects that derive from the area of surfaces.

One example is the notion of \textit{quantum expansion}. Recall, the classical expansion of a surface $\sigma$ at a point $y \in \sigma$ is defined as the trace of the null extrinsic curvature at $y$. Equivalently, one can define the classical expansion as a functional derivative,
\begin{equation}
    \theta[\sigma; y] = \frac{1}{\sqrt{h(y)}}\frac{\delta A[V]}{\delta V(y)}~,
\end{equation}
where $h$ represents the area element of the metric restricted to $\sigma$, inserted to ensure that the functional derivative is taken per unit proper area, not coordinate area. The function $V(y)$ is used to specify the affine location of $\sigma$ and nearby surfaces along a congruence of null geodesics orthogonal to $\sigma$.
The above definition of the classical expansion is needlessly complicated, in that it invokes the entire surface $\sigma$, even though $\theta$ depends only on its local extrinsic curvature at $y$. However, this definition naturally generalizes to the quantum expansion, $\Theta$, which does depend on all of $\sigma$:
\begin{equation}
  \Theta[\sigma; y] \equiv \frac{4G\hbar}{\sqrt{h(y)}}
  \frac{\delta S_{\text{gen}}[V]}{\delta V(y)}~.
\end{equation}
  
As in the classical case, we can use the notion of expansion to define certain types of surfaces (see Sec.~\ref{sec-cnd}). Let $\Theta_\pm$ be the quantum expansion of the future-directed light-rays orthogonal to a surface $\mu_Q$. (As before, we take the $+$ label to refer to the direction of spatial infinity.) If $\Theta_+ \leq 0$ ($\Theta_+ = 0$) and $\Theta_- \leq 0$, then we call $\mu_Q$ a \textit{quantum (marginally) trapped surface}.

Quantum trapped surfaces, in the semiclassical setting, have some of the properties obeyed by trapped surfaces in the classical setting. For example, trapped surfaces cannot lie outside the black hole, assuming weak cosmic censorship and the Null Energy Condition. When the NEC is violated, they can; however, quantum trapped surfaces must still lie inside or on the horizon~\cite{Wal13} (still assuming weak cosmic censorship). This will prove to be important for our formulation of the quantum Penrose inequality.

A \textit{quantum future holographic screen}, or Q-screen, is a hypersurface foliated by quantum marginally trapped surfaces. Assuming the quantum focussing conjecture~\cite{BouFis15a}, Q-screens obey a Generalized Second Law~\cite{BouEng15c}.

\subsection{Lessons From the Counterexample}
\label{sec-motivation}

The failure of the classical PI in the presence of quantum matter (Sec.~\ref{sec-boulware}) illustrates the need for a Quantum Penrose Inequality. It also motivates some of the choices we will make below.

Let us distinguish two different time-scales: the time for the negative energy of the Boulware-like state to enter the black hole, and the evaporation time. The former is of order the scrambling time $\Delta t_s\sim R \log (R/l_P)$. The latter is much greater, of order $R^3/G\hbar$.

On the shorter time-scale, the process results in an outcome very similar to that invoked in motivating the classical Penrose inequality: a Kerr black hole with area $A_{\rm late}$ and no further evolution. That is, we neglect evaporation since it occurs on a much greater timescale; and by construction, no matter that will ever enter the black hole. Thus, the mass should obey $16\pi G^2m^2\geq A_{\rm late}$.

The key difference to the classical case is that the ``late'' area need not be greater than the area of trapped surfaces at earlier times; indeed our counterexample shows that it will not be. However, we know that the Generalized Second Law (GSL) takes the place of the area theorem in this setting. Thus, we expect that the generalized entropy of earlier quantum trapped surfaces should be less than $A_{\rm late}/4G\hbar$. And so, the generalized entropy of quantum trapped surfaces should replace the area of trapped surfaces when we replace the classical by a Quantum Penrose Inequality.

This argument is based on the GSL for the event horizon, and so involves an intermediate step where one argues that the generalized entropy of a quantum marginally trapped surfaces inside the black hole will not be greater than that of the event horizon. To avoid this step, we can generalize the second heuristic argument for the classical Penrose inequality, which was based on the area theorem for future holographic screens. Q-screens obey a GSL that interpolates directly between different marginally quantum trapped surfaces. If a suitable Q-screen connects $\mu_Q$ to the late-time event horizon, this establishes a Quantum Penrose Inequality. Of course this is far from a trivial assumption; our goal here was only to gain some intuition.

In the above heuristic arguments, it was important that the late-time generalized entropy should be given just by $A_{\rm late}$, i.e., that no entropy remains outside of the black hole. However, this will not be the case in general examples. This will motivate our choice, below, that the generalized entropy entering the Quantum Penrose Inequality should be evaluated on slices that remain inside the black hole. We will discuss this important issue further in Sec.~\ref{sec-wrong1}.

\subsection{Formulation}
\label{sec-formulation}

We will now obtain a Quantum Penrose Inequality from the classical PI, in three steps. First, we replace the area with generalized entropy in Eq.~(\ref{eq-pi}):
\begin{equation}
A \to 4G\hbar\, S_{\rm gen} \equiv A+4G\hbar\, S_{\rm out}~.
\end{equation}
Thus we propose an inequality of the form
\begin{equation}
  m\geq \sqrt{\frac{\hbar S_{\rm gen}}{4\pi G}}~.
  \label{eq-qpiinc}
\end{equation}

Secondly, we must specify the surfaces to which the inequality can be
applied. In the classical case, a surface $\mu$ has to be trapped for the Penrose inequality to apply, corresponding to criteria satisfied by the classical expansion. For the QPI, it is natural to apply the same criteria to the quantum expansion:
\begin{equation}
  \theta \to \Theta~.
\end{equation}
Thus in Eq.~(\ref{eq-qpiinc}), $S_{\rm gen}$ is the generalized entropy of any surface $\mu_Q$ that is quantum trapped. We expect that the most interesting bounds will obtain when $\mu_Q$ is quantum marginally trapped, and we will only consider this case in all examples below.

Next, we must specify on which achronal hypersurface the generalized entropy appearing in Eq.~(\ref{eq-qpiinc}) should be computed. As we will explain in Sec.~\ref{sec-wrong1}, this {\em cannot\/} be chosen to be a Cauchy surface of the outer wedge.
Instead, we will propose that this hypersurface should be entirely contained in the ``black hole region'' $B\equiv M-J^-({\cal I}^+)$, i.e., inside or on the horizon.

\begin{figure}%
\centering
    \includegraphics[width=.475\textwidth]{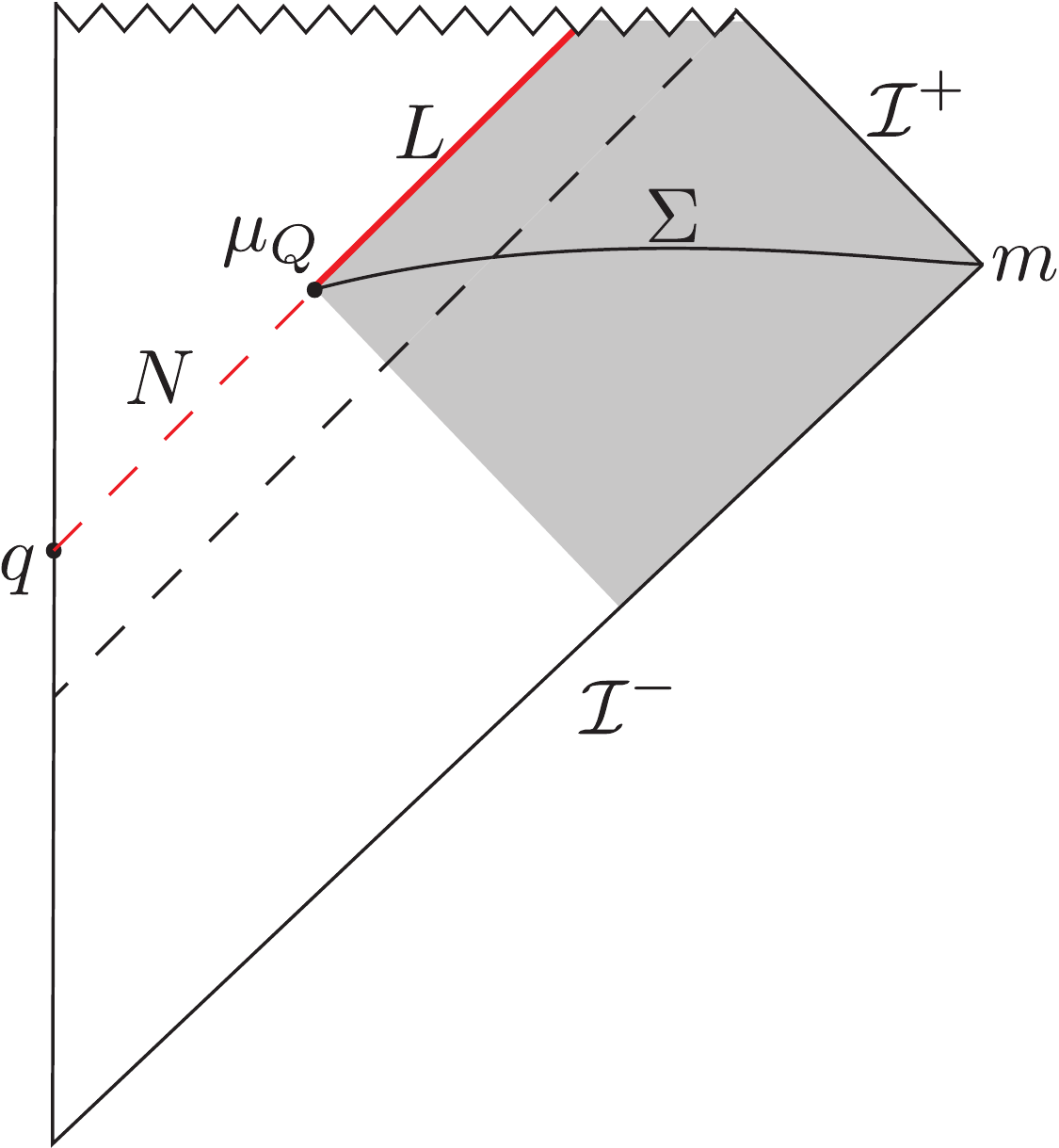}
    \caption{The Quantum Penrose Inequality bounds the mass at infinity in terms of the generalized entropy of a quantum marginally trapped surface $\mu_Q$. The generalized entropy must be evaluated on the lightsheet $L$ (red line), {\em not} on a Cauchy surface $\Sigma$ of the outer wedge $O_{W}[\mu_{Q}]$ (shaded region).}%
    \label{fig-slices43}%
  \end{figure}
More precisely, we require that $S_{\rm gen}$ should be evaluated on the ``future portion'' of the boundary of the outer wedge,
\begin{equation}
L(\mu_Q) \equiv \dot O_W(\mu_Q)- I^-(O_W(\mu_Q))~.
\end{equation}
See Fig.~\ref{fig-slices43}. $L$ is generated by the congruence of future-directed outgoing null geodesics orthogonal to $\mu_Q$~\cite{Wald,AkeBou17}. Their initial quantum expansion is $\Theta_+=0$ by construction, so assuming the QFC~\cite{BouFis15a}, $\Theta_+\leq 0$ everywhere on $L$. Hence $L$ will be a (quantum) lightsheet of $\mu_Q$. Assuming an appropriate version of weak cosmic censorship, $L$ will terminate on the singularity inside the black hole. (Strictly, in order to remain in the semi-classical regime, one should terminate $L$ slightly earlier, resulting in a second area term that can be made small by approaching the singularity.)

Note that the surface $\mu_{Q}$ must be quantum trapped with respect to $L$; it need not be quantum trapped with respect to any other hypersurface, such as a Cauchy surface of $O_W(\mu_Q)$. To find a suitable $\mu_Q$, consider a null hypersurface $N$ inside the black hole, for example the boundary of the future of an event $q$ inside the black hole; see Fig.~\ref{fig-slices43}. Typically the area of $N$ will increase near $q$ and later decrease towards the singularity. Hence the area will have a maximum on some cut of $N$, and the generalized entropy of cuts of $N$ (computed with respect to the future of the cuts on $N$) will have a maximum on some nearby cut. This cut will be a suitable quantum marginally trapped surface $\mu_Q$, and later cuts will also be quantum trapped. 

Finally, we must impose a requirement analogous to the minimum area condition imposed on $\mu$ in the classical case. This condition demanded that there exist a Cauchy surface of $O_W$ on which no surface enclosing $\mu_Q$ has area less than $\mu_Q$. Here, we will instead consider the generalized entropy of any surface $\nu$ enclosing $\mu_Q$, computed on the boundary of the future of the outer wedge of $\nu$. For the QPI to apply to a quantum trapped surface $\mu_Q$, we demand that there exist a Cauchy surface of $O_W[\mu_Q]$ on which no enclosing surface $\nu$ satisfies $S_{\rm gen}[\dot O_W(\nu)- I^-(O_W(\nu))]<S_{\rm gen}[L(\mu_Q)]$.

To summarize, we propose that the mass at spatial infinity of an asymptotically flat spacetime satisfies the Quantum Penrose Inequality
\begin{equation}
  m\geq \sqrt{\frac{\hbar S_{\rm gen}[L(\mu_Q)]}{4\pi G}}~,
  \label{eq-qpi}
\end{equation}
where $S_{\rm gen}$ is computed on the future-outgoing lightsheet of $\mu_Q$, and $\mu_Q$ is any quantum trapped surface homologous to spatial infinity that has minimal generalized entropy on some Cauchy surface of its outer wedge, in the sense described above.

We close by discussing a subtlety that introduces a small uncertainty in the formulation of the QPI. In Eq.~(\ref{eq-qpi}), we used the classical functional relation between the area and mass of Schwarzschild black holes; we merely replaced the area with the generalized entropy. In fact, there will be a field-content-dependent quantum correction to the functional relation itself. However, this correction is small compared to the difference between our QPI and the classical Penrose inequality.

This is easier to discuss in asymptotically Anti-de Sitter (AdS) space, where the Schwarzschild black hole can be in thermal equilibrium. Therefore, we will revisit the issue in more detail in Sec.~\ref{sec-ads}. In general, the black hole exterior will have nonzero energy density in equilibrium. This is a kind of Casimir energy associated with the potential well provided by the near horizon zone. It contributes to the total mass at infinity; but since it stays outside the black hole, it will not contribute to $S_{\rm gen}[L(\mu_Q)]$.

By dimensional analysis, one expects each field theory degree of freedom to contribute an amount of order $\hbar/R$ to this Casimir energy. In Eq.~(\ref{eq-qpi}), this is equivalent to changing the area or generalized entropy by $O(c)$, where $c$ is the number of matter quantum fields. For large black holes in AdS, it is possible to determine this correction and include it in the QPI (see Sec.~\ref{sec-ads}). In general, however, we are presently unable to determine it.

Since $S_{\rm gen}$ is $O(\hbar^{-1})$ and $c$ is $O(1)$, the undetermined Casimir term in Eq.~(\ref{eq-qpi}) is subleading. But naively, it is comparable to the refinement we introduced in passing from the classical Penrose inequality to the QPI. However, the Casimir correction cannot be enhanced by factors proportional to $\hbar^{-1}$. Thus it is much smaller than the violations of the classical Penrose inequality that were exhibited in Sec.~\ref{sec-boulware}. Because of the $\hbar^{-1}$  enhancement, Eq.~(\ref{eq-pi}) can be violated by a {\em classical} amount through quantum effects. Correspondingly, a successful QPI cannot be a small modification of the classical Penrose inequality. Indeed, it is not: as we shall demonstrate in the next section, the counterexample to Eq.~(\ref{eq-pi}) is evaded by Eq.~(\ref{eq-qpi}). In this and many other interesting examples, the Casimir correction is small compared to the difference between Eq.~(\ref{eq-pi}) and Eq.~(\ref{eq-qpi}).

\section{Evidence for the Quantum Penrose Inequality}
\label{sec-evidence}

We will now analyze the validity of our proposal in a number of examples. In the process, we will gain some intuition about the key quantity that appears in it: $S_{\rm gen}[L]$, the generalized entropy of the future-outgoing lightsheet $L$ of a quantum marginally trapped surface $\mu_Q$. 

\subsection{Black Hole in the Unruh State}

\begin{figure}
\centering
    \includegraphics[width=.5\textwidth]{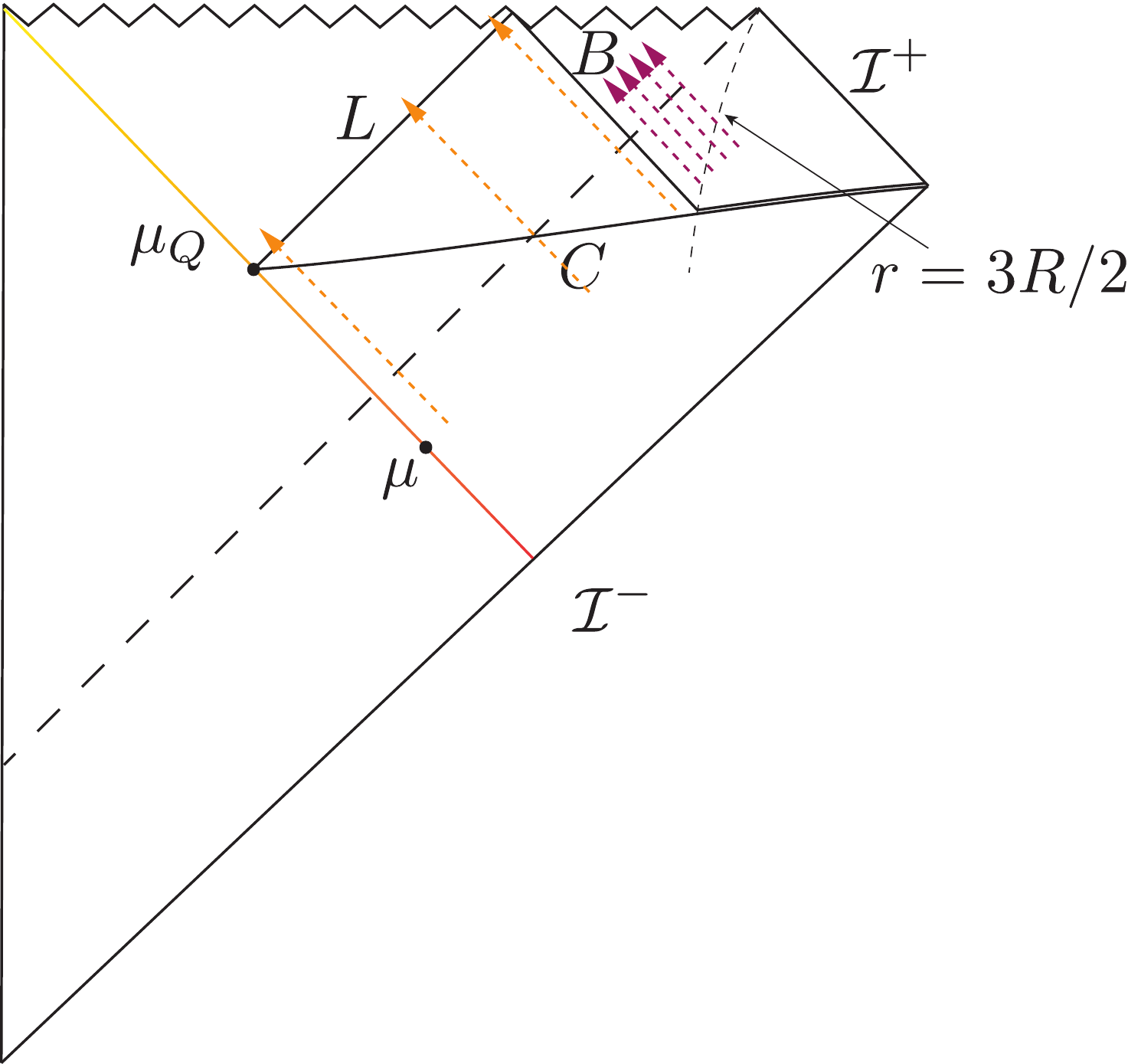}
    \caption{Black hole formed from the collapse of a null shell (orange line). The classically marginally trapped surface $\mu$ lies a Planckian distance outside of the event horizon. The quantum marginally trapped surface $\mu_{Q}$ lies a Planckian distance inside the horizon. The lightsheet $L(\mu_Q)$ captures $\sim \log (R/l_P)$ infalling Hawking modes (orange dashed lines); in the Unruh states these modes are unoccupied and so contribute negative entropy on $L$, compared to the Hartle-Hawking state. $L$ ends at the singularity and does not encounter any later infalling modes (purple dashed lines). The entropy on $L$ can also be computed using the mutual information, $S_L = S_C-S_B+I(L:B)$.}
\label{fig-ex1}
\end{figure}

As a first example, consider a black hole formed from collapse of a null shell; see Fig.~\ref{fig-ex1}. This is the example we analyzed in the context of the classical Penrose inequality, at the beginning of Sec.~\ref{sec-boulware}. We showed there that the CPI is saturated, since the area of the classically marginally trapped surface $\mu$ immediately after the collapse satisfies
\begin{equation}
  16\pi G^2 m^2=A[\mu]~.
  \label{eq-e11}
\end{equation}
Here we are interested in a quantum marginally trapped surface with largest generalized entropy, for which the QPI provides the greatest lower bound on the mass. The area of (quantum) trapped surfaces decreases along with the event horizon, and the contribution from the entropy term is approximately time-independent. Hence we will again choose the earliest possible surface $\mu_Q$, right after the collapse.

The quantum marginally trapped surface $\mu_Q$ must lie inside the event horizon~\cite{Wal13}, whereas $\mu$ lies outside. Therefore
\begin{equation}
  A[\mu_Q]<A[\mu]~.
  \label{eq-e12}
\end{equation}

We now turn to estimating $S_{\rm gen}[L]$. Strictly, $S_{\rm gen}[L]$ should be computed from the quantum state on a global Cauchy surface $\Sigma$ that contains $L$. One would first compute the (divergent) field theory entropy $S[L]$ by tracing over the complement of $L$ on $\Sigma$. One would then add the gravitational counterterms whose leading contribution is $A[\mu_Q]$. Locally, in a vacuum state, one expects $S_{\rm gen}\approx A[\mu_Q]/4G\hbar$, where $G$ is the ``infrared'' value of Newton's constant that would be observed at large distances.

However, the state on $L$ is not a standard vacuum state. $L$ nearly coincides with the black hole horizon for a time $t\ll \Delta t_s$, where $\Delta t_s$ is the scrambling time. The vacuum state on the horizon is the Hartle-Hawking state, which contains ingoing radiation. The ingoing radiation on $L$ is entangled with modes on the other side of $L$. This contribution must be canceled by the counterterm so as to obtain $S_{\rm gen}\approx A[\mu_Q]/4G\hbar$ in the Hartle-Hawking state.

The actual state we consider here is the Unruh state, which does not have this ingoing radiation. As a result, the lightsheet will contain less entropy than in the vacuum state. Thus
\begin{equation}
  S_{\rm gen}[L]<\frac{A[\mu_Q]}{4G\hbar}~.
  \label{eq-e13}
\end{equation}
Combined with Eqs.~(\ref{eq-e11}) and (\ref{eq-e12}) this establishes that the QPI is satisfied (and not saturated) in this example.

We would like to go further and estimate the ``gap'' by which the QPI fails to be saturated in this example,
\begin{equation}
  \Delta \equiv \frac{4\pi G}{\hbar} m^2- S_{\rm gen}[L]~.
\end{equation}
We will be interested only in the order of magnitude of this gap and so will make a number of approximations. We refer to Sec.~\ref{sec-boulware} for notation and conventions.

First, we will assume that the higher angular momentum modes, $\ell>0$, in the near-horizon zone completely reflect off of the angular momentum barrier and so will behave as if they were in the Hartle-Hawking state. In this approximation, the Unruh state differs only through the spherical ($\ell=0$) modes, which we treat as having no angular momentum barrier at all. We also assume that the ingoing and outgoing s-waves do not interact.

A Planck sized, radially outgoing wavepacket starting a Planck distance from the horizon will be redshifted in such a way that its proper distance from the horizon remains comparable to its proper wavelength, while it propagates in the near horizon zone, $r\lesssim 3R/2$. Thus, the number of independent ingoing s-wave modes captured by $L$ is of order $\log (R/l_P)$, as shown in Fig.~\ref{fig-ex1}. In other words, $L$ ``sees'' what enters the black hole in the first scrambling time after infalling geodesics that would have crossed $\mu_Q$ (see also Appendix~\ref{sec-qts}).

Every such mode would contribute $O(1)$ entropy in the Hartle-Hawking state but is pure in the Unruh state (since it is in the ground state). The missing entropy, and the gap to saturating the QPI, is thus
\begin{equation}
  \Delta \sim \log\frac{R}{l_P}~.
  \label{eq-e14}
\end{equation}

The entropy on null surfaces can have surprising and counterintuitive properties~\cite{BouCas14b}. As a check on the above arguments, we now verify this result by evaluating $S_{\rm gen}[L]$ using an alternative method, in which von Neumann entropies are evaluated only on spacelike hypersurfaces.\footnote{We thank Aron Wall for suggesting this approach.}

The mutual information of any two systems is defined in terms of the von Neumann entropies of the individual and joint systems as follows:
\begin{equation}
  I(L:B)\equiv S_L+S_B-S_{LB}~.
\end{equation}
Here we consider the lightsheet $L$ and the partial Cauchy surface $B$ shown in Fig.~\ref{fig-ex1}. We take $B$ to be null until it meets the end of the near horizon zone, $r=3R/2$, and to coincide approximately with a constant $t$ hypersurface outside of this radius. To stay in the semiclassical regime, one can terminate $L$ slightly before the singularity. We can choose this terminal surface to have area $c l_P^2$, where $1\ll c\ll \log (R/l_P)$. The second inequality ensures that its contribution will be subleading to our result.

Note that the joint system $LB$ is equivalent by unitary evolution to the purely spacelike Cauchy surface $C$. We can thus evaluate the von Neumann entropy on $L$ as
\begin{equation}
  S_L = S_C-S_B+I(L:B)~.
  \label{eq-slmi}
\end{equation}
Moreover, $L$ and $C$ have the same boundary, $\mu_Q$, whereas $B$ has a boundary of negligible area. It follows that
\begin{equation}
S_{\rm gen}[L]= S_{\rm gen}[C]-S_B+I(L:B)~.
\end{equation}
We chose $\mu_Q$ to be just after black hole formation, so there will be no outgoing Hawking radiation present on $C$. In the Unruh state, the ingoing spherical modes in the near-horizon zone are unoccupied, which reduces the entropy by $\log (R/l_P)$ compared to the Hartle-Hawking value. Hence
\begin{equation}
  S_{\rm gen}[C]- \frac{A[\mu_Q]}{4G\hbar}\sim \log\frac{R}{l_P}~.
\end{equation}
In our approximation, $B$ captures the same outgoing modes as $C$, but none of the ingoing modes that cross $L$, so $S_B = 0$. There is no data on $L$ that is entangled with data on $B$, so $I(L:B) = 0$. Hence Eq.~(\ref{eq-slmi}) implies $S_{\rm gen}[L]=S_{\rm gen}[C]$ in our example. Since $16\pi G^2m^2=A[\mu] = A[\mu_Q]+O(l_P^2)$, we recover Eq.~(\ref{eq-e14}).

Note that the Planck length enters Eq.~(\ref{eq-e14}) through the position of the quantum marginally trapped surface $\mu_Q$, which is a proper distance of order $l_P$ inside of the event horizon (or of $\mu$). It would appear, therefore, that $\Delta$ could be minimized if one could arrange for $\mu_Q$ to lie a distance comparable to $R$ inside the horizon. However, this requires a large perturbation of the black hole, to which the current analysis does not apply. We will revisit this question in Sec.~\ref{sec-qpigsl}.

\subsection{Near-Saturation of the QPI}

In the previous subsection, we found that in a newly formed Schwarzschild black hole with no exterior matter, the QPI will be satisfied but not quite saturated, with a gap of $\Delta\sim \log (R/l_P)$. The gap is only logarithmic, but it still becomes arbitrarily large for large black holes. Here we show that the logarithmic gap can be eliminated. Thus, the QPI can be saturated up to a fixed gap of order a Planck area, which we do not have full control over.

The simplest way to accomplish this is to time-reverse the state of the semiclassical fields on the partial Cauchy surface $C$ shown in Fig.~\ref{fig-ex1}. In our approximation, this will not affect the $\ell>0$ modes, but it will put the spherical waves in a time-reversed Unruh state. That is, the outgoing modes will be unoccupied and the ingoing modes will be occupied, reversing the situation considered in the previous subsection. Crucially, this modification will not change the mass $m$ at infinity, so we still have
\begin{equation}
  16\pi G^2 m^2 = A[\mu] = A[\mu_Q]+O(l_P^2)~.
\end{equation}

Because of the restriction to semiclassical modes, there is a cutoff near $\mu_Q$ at least of order $l_P$. Thus, while the initial conditions we now impose are somewhat unnatural, they will persist only for one scrambling time $\Delta t_s \sim R\log (R/l_P)$. After this time, the black hole will begin to evaporate. In particular, unlike the full Boulware state, there is no singularity at the horizon. Note also that this state differs from the one we considered in Sec.~\ref{sec-boulware} in that the $\ell>0$ modes are not in the Boulware vacuum.

The lightsheet $L$ is sensitive only to the ingoing part of the radiation, so its generalized entropy will be the same as it would be in the Hartle-Hawking state:
\begin{equation}
  S_{\rm gen}[L] = \frac{A[\mu_Q]}{4G\hbar}~.
\end{equation}
Thus we find that the QPI is nearly saturated:
\begin{equation}
  \Delta \equiv \frac{4\pi G}{\hbar} m^2- S_{\rm gen}[L]\sim O(1)~.
\end{equation}

\subsection{Perturbative Regime: QPI from the GSL}
\label{sec-qpigsl}

Next, we will consider the more general case where matter enters into the black hole after its formation. We consider the same formation process as above. We will again focus on $\mu_Q$ right after formation so as to obtain the tightest bound. But now we will allow for a nontrivial quantum state outside of the black hole. This could be an ordinary matter system carrying some thermodynamic entropy. It could also be a quantum state with negative energy, such as the Boulware-like state that we considered in Sec.~\ref{sec-boulware} as a counterexample to the CPI.

The future-outgoing lightsheet $L$ of $\mu_Q$ will only receive matter that falls into the black hole within the first scrambling time after $\mu_Q$; see Fig.~\ref{fig-ex1}. To be precise, consider a family of radially infalling geodesics that are initially at rest at some large radius $r\gg R$. The geodesics are all at the same angle but shifted in time. It is easy to check that the geodesic that passes through $\mu_Q$ and the last geodesic that reaches $L$ are separated at large radius by a time of order $\Delta t_s \sim R\log (R/l_P)$. Any matter that falls in later will hit the singularity before reaching $\Sigma$. This statement does not depend on the initial radius, and it also holds also for ingoing null geodesics; see Appendix ~\ref{sec-qts}.

In the following subsection, we will consider the effects of matter that falls in after the first scrambling time and so does not reach $L$. However, now we will focus on matter that can be  registered on $L$. By the above argument, we can take this matter to reside within the near-horizon zone, $R<r<3R/2$, on the partial Cauchy surface $C$. Let $H$ be the portion of the event horizon to the future of $C$, and let $S_{\rm gen}[H]$ be its generalized entropy.

We begin by making a simplifying assumption that will be relaxed below, that all of the matter that falls across the horizon will also cross $L$ (as opposed to passing through the portion of $B$ inside the black hole). The quantum marginally trapped surface $\mu_Q$ and the boundary of $H$ have approximately the same area, so there is a simple relationship between the entropy on $H$ and $L$: 
\begin{equation}
S_{\rm gen}[L]=S_{\rm gen}[H] - \Delta S[H_{\rm late}] + \mathcal{O}(1),
\label{eq-HLrelation}
\end{equation}
where $H_{\rm late}$ is the portion of the horizon above a sufficiently late Cauchy slice, when the black hole has relaxed to equilibrium, but early enough that negligible Hawking radiation has been produced.

We have assumed a state in which there is negligible mutual information between $L$ and $H_{\rm late}$. For example, if the black hole simply evaporates with no further matter falling in, $\Delta S[H_{\rm late}]$ is the (negative) renormalized entropy that exists on the horizon in the Unruh state (due to the missing infalling modes when to compared the Hartle-Hawking state).

From
\begin{equation}
S_{\rm gen} [H_{\rm late}] - \Delta S[H_{\rm late}]=\frac{A_{\rm late}}{4G\hbar}
\label{eq-HHrelation}
\end{equation}
and Eq.~\eqref{eq-HLrelation}, the QPI follows:
\begin{equation}
  S_{\rm gen}[L]=S_{\rm gen}[H] - \Delta S[H_{\rm late}] \leq S_{\rm gen}[H_{\rm late}] - \Delta S[H_{\rm late}]
 =\frac{A_{\rm late}}{4G\hbar} \leq
  \frac{4\pi G}{\hbar} m^2~.
\end{equation}
The first inequality in this sequence is the GSL for event horizons. Note that we have ignored the $\mathcal{O}(1)$ additive uncertainty in Eq.~\eqref{eq-HLrelation} in light of the discussion at the end of Sec.~\ref{sec-QPI}.

This argument establishes the QPI for a large class of examples, including the Boulware-like state that served as a counterexample to the classical Penrose inequality in Sec.~\ref{sec-boulware}. In this case, $A_{\rm late}$ (which sets the mass) will be significantly smaller than the area of the trapped surface $\mu$. Here we use the quantum trapped surface $\mu_Q$, but its area is almost the same as that of $\mu$. What saves the QPI is the contribution of the entropy on $L$, which is negative in this example. Specifically, the GSL guarantees that the lower bound, $S_{\rm gen}[L]$, is smaller than the area of $\mu_Q$ by a sufficient amount for the QPI to hold.

In the case where positive entropy registers on $H$ and $L$, our QPI is stronger than the classical Penrose inequality. The lightsheet ``knows'' that more matter will enter the the black hole after $\mu_Q$, and the GSL ``knows'' that this will result in an area increase. Effectively, this larger area becomes the lower bound on the mass.

\subsection{Failed Counterexample: Negative Energy That Misses the Lightsheet}

In the previous subsection, we considered the case where all matter outside the quantum trapped surface $\mu_Q$ crosses its lightsheet $L$. Here we generalize to discuss matter for which this does not happen. In this case, we cannot use the GSL for the event horizon to constrain the relation between $S_{\rm gen}[L]$ and the mass at infinity. However, we will give some plausibility arguments for the validity of the QPI. 

In the previous subsections, we argued that the QPI will hold true if all matter outside of $\mu_Q$ passes through $L$. We can think of the present situation as a complication where we add matter that does not satisfy this property. Since this cannot affect $S[L]$, the only way that the QPI can now be violated is if the matter we added contributes negative mass at infinity. We will now argue that this is impossible in the semiclassical regime.

\begin{figure}
\centering
    \includegraphics[width=1\textwidth]{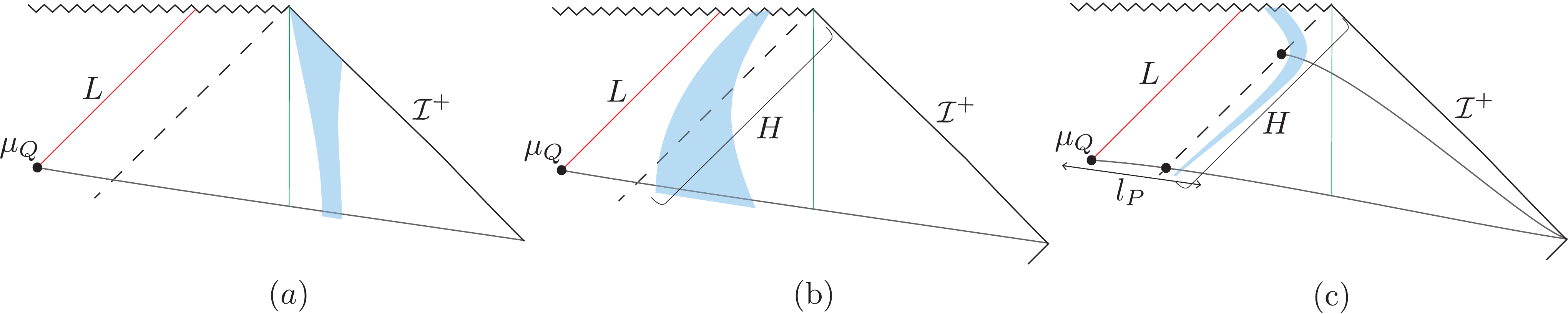}
    \caption{The QPI is threatened by any negative energy (blue worldvolume) that fails to register on the lightsheet $L$. We analyze three possibilities but find that none of them leads to a violation of the QPI. (a) Negative energy outside of the near horizon zone (vertical green line). (b) Negative energy that enters the black hole soon after $\mu_Q$ but evades $L$ by accelerating outward. (c) Negative energy that remains near the black hole for more than a scrambling time.}%
    \label{fig-thethree}%
\end{figure}

Matter outside of $\mu_Q$ can fail to register on $L$ for any of the following three reasons (see Fig.~\ref{fig-thethree}):
\begin{enumerate}
\item The matter never enters the black hole.
\item The matter enters the black hole during the first scrambling time after $C$ but escapes through the portion of $B$ inside the black hole.
\item The matter enters the black hole later than a scrambling time after $C$.
\end{enumerate}

In the first case, the matter can be approximately treated as isolated from the black hole. But the total mass of isolated systems is positive, so distant systems can never cause violations of the QPI. (This does not rule out regions with negative energy, but it implies that sufficient positive energy must be present nearby.)

In the second case, the matter system can be initially near the black hole and so could have regions of negative energy density (as in the example of Sec.~\ref{sec-boulware}). However, in order to miss $L$, it would have to accelerate outwards after crossing the horizon. This requires positive energy. We will not attempt to demonstrate here that this always results in a net positive mass contribution; our goal is only to note that the QPI is not obviously violated in this setup. This question merits further study.

In the third case, we again must choose the matter system to be close to the horizon if we wish to give it negative energy. For example, the Boulware-like state of Sec.~\ref{sec-boulware} would qualify. However, by assumption this state would have to be present more than one scrambling time after $C$. Moreover, the modes for which it is possible to obtain net negative energy are those that make up the thermal atmosphere of the black hole; these modes evolve exponentially close to the horizon under backward time evolution. Thus the state on $C$ would contain transplanckian energy density (similar to a firewall). The initial state would not be a semiclassical state. This argument is robust and rules out an entire class of what naively seemed like promising counterexamples. We view this as nontrivial evidence in favor of our proposal.

\section{Alternative Proposals}
\label{sec-alternatives}

In this section we consider various alternative conjectures for the QPI. In Sec.~\ref{sec-no} we give counterexamples to proposals that might otherwise seem natural. In Sec.~\ref{sec-maybe} we discuss modifications of our proposal that appear viable, and we explain why we are not currently advocating for them. 

\subsection{Nonviable Alternatives}
\label{sec-no}

We will now discuss several alternative conjectures for a QPI that we considered in the process of this work. Our goal is to explain our choice in Sec.~\ref{sec-QPI}, and to illustrate that the problem is rather constrained. This proves neither that our formulation is unique, nor that it is correct. But we will see that it is remarkably difficult to find any alternative statement of the QPI that is not immediately ruled out. 

\paragraph*{Cauchy surfaces that reach spatial infinity}
\label{sec-wrong1}

\begin{figure}%
\centering
    \includegraphics[width=.8\textwidth]{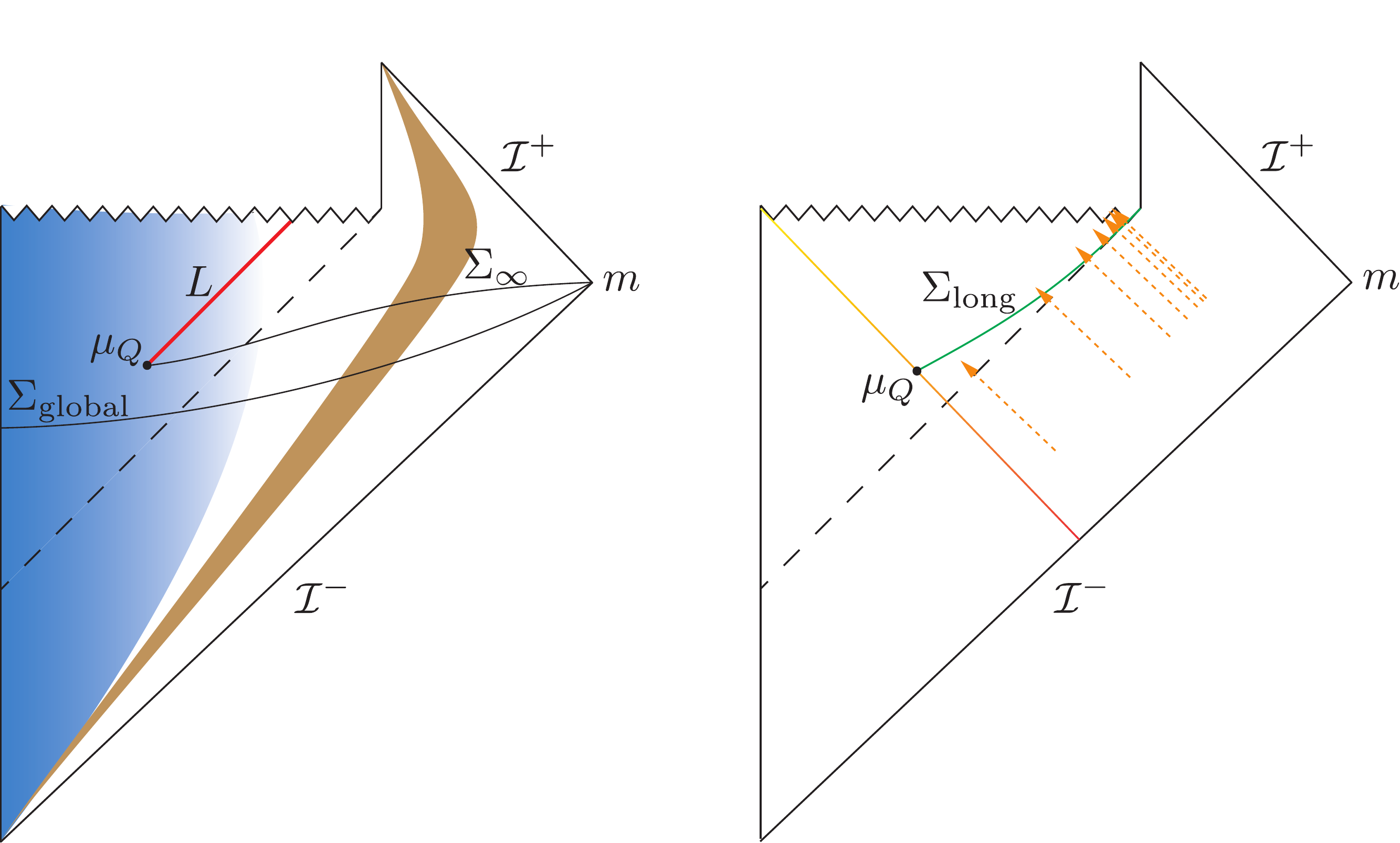}
    \caption{Left: the generalized entropy on the slice $\Sigma_{\infty}$ can be dominated by distant soft particles (brown) and so does not yield a viable lower bound on the mass. The global Cauchy surface $\Sigma_{\text{global}}$ plays a role in an alternative proposal discussed in the main text. Right: the long slice $\Sigma_{\rm long}$ captures all of the missing infalling Hawking modes.}%
    \label{fig-slices}%
\end{figure}

First, we explain why we do not allow $\Sigma[\mu_Q]$ to reach outside the black hole. This prohibition is motivated by the asymptotically flat case, to which we will specialize for now. Let $\Sigma_\infty$ be a Cauchy surface of $O_W[\mu_Q]$, in violation of our requirements. An example is the black slice in the Fig.~\ref{fig-slices}. Let $S_{\rm gen}[\Sigma_\infty(\mu_Q)] $ be the generalized entropy evaluated on $\Sigma_\infty$. The alternative QPI thus would take the form
\begin{equation}
  m \stackrel{?}{\geq} \sqrt{\frac{\hbar}{4\pi G} S_{\rm gen}[\Sigma_\infty(\mu_Q)]}~.
  \label{eq-wrong1}
\end{equation}

But it is easy to find a counterexample to Eq.~(\ref{eq-wrong1}): an arbitrary amount of matter entropy can be placed in regions far from the black hole, at arbitrarily little cost in mass. We now discuss this in detail.

Consider a dilute gas of $N$ photon wave packets, each of characteristic size $\lambda$. Each photon occupies a region of volume $\lambda^3$, so the photons can be dilute if they occupy a region of volume $N\lambda^3$.  We can take each photon to be in a mixed state (say, of polarizations), and in a product state with respect to the rest of the universe. Then the gas contributes of order $N$ to the generalized entropy on $\Sigma$.

We take the gas to be very far from the black hole or any other matter, so that gravitational binding energy to other objects is negligible. The gravitational binding energy of the photon cloud itself will be negligible if $N G\hbar/\lambda\ll N^{1/3} \lambda$, so we shall take $\lambda\gg N^{1/3} l_P$, where $l_P\equiv (G\hbar)^{1/2}$ is the Planck length. Then the gas of photons contributes a mass of order $N\hbar/\lambda$ to the ADM mass. This mass contribution can be taken to be arbitrarily small by taking $\lambda\to \infty$ at fixed $N$ without violating any of the previous assumptions.

We are still free to choose $N$ to take any value we like. Thus we have found a family of initial data with bounded $m$ but unbounded $S_{\rm gen}[\mu_Q]\approx c_1+c_2N$, where $c_1$ and $c_2$ are independent of $N$. For large enough $N$, this leads to a violation of Eq.~(\ref{eq-wrong1}).

\paragraph*{Area of marginally quantum trapped surfaces}

A second alternative conjecture would be to use only the area of $\mu_Q$, not its generalized entropy:
\begin{equation}
  m \stackrel{?}{\geq} \sqrt{\frac{A[\mu_Q]}{16\pi G^2}}~.
  \label{eq-wrong2}
\end{equation}
That is, one would conjecture that  Eq.~(\ref{eq-pi}) holds if $A$ is taken to be the area of a quantum trapped surface. This possibility is attractive because the entropy of distant soft radiation would never contribute to the lower bound in the first place.

However, Eq.~(\ref{eq-wrong2}) is ruled out (among other reasons) by the Boulware-like  counterexample to the classical Penrose inequality. This is because the area of the bifurcation surface will receive only a correction that can be made parametrically small. This follows from the remarks concerning the classically marginally trapped surface at the end of Sec.~\ref{sec-boulware}. The same argument implies that the marginally quantum trapped surface area receives only a parametrically small correction, which cannot compete with the large decrease in mass.

\paragraph*{Subtracting global entropy; interior generalized entropy}

Let us revisit the proposal of Sec.~\ref{sec-wrong1} and consider the generalized entropy $S_{\rm gen}[\Sigma_\infty(\mu_Q)]$ of a marginally trapped surface $\mu_Q$, evaluated on a Cauchy surface that reaches outside of the black hole all the way to spatial infinity. This proposal suffered from the problem that distant soft modes can contribute unbounded entropy with bounded energy, so $S_{\rm gen}[\Sigma_\infty(\mu_Q)]$ is unrelated to any lower bound on the mass.

A natural idea is to subtract the von Neumann entropy on a global Cauchy surface (see Fig.~\ref{fig-slices}):
\begin{equation}
  m \stackrel{?}{\geq} \sqrt{\frac{\hbar (S_{\rm gen}[\Sigma_\infty(\mu_Q)]
      -S[\Sigma_{\rm global}]}{4\pi G}}~.
  \label{eq-wrong3}
\end{equation}
If the distant soft modes have the same entropy in the global state as in the generalized entropy, then their dangerous contribution will cancel out.

However, this need not be the case. Consider a collapsing star that forms a Schwarzschild black hole of area $A$. The entropy of the star can be of order $S_{\rm star}\sim (A/G\hbar)^{3/4}$ or even $S_{\rm star}\sim A/G\hbar$~\cite{BouFre10a}. We can chose the global state to contain only distant soft radiation that purifies the star, so that $S[\Sigma_{\rm global}]=0$ and
\begin{equation}
  m=\sqrt{\frac{A[\mu_Q]}{16\pi G^2}}+\epsilon~,
\end{equation}
where $\epsilon$ can be arbitrarily small. But then
\begin{equation}
  S_{\rm gen}[\Sigma_\infty(\mu_Q)]\approx \frac{A[\mu_Q]}{4G\hbar}+ S_{\rm star}~,
\end{equation}
so that Eq.~(\ref{eq-wrong3}) is violated.

The violation in our example remains bounded, since $S_{\rm star}$ cannot exceed $A[\mu_Q]/4G\hbar$ by the GSL. One might consider absorbing this violation by adding a correction factor of $1/2$ to the right hand side of Eq.~(\ref{eq-wrong3}).  But by considering initial data with a second asymptotic region, one can arrange $S[\Sigma_{\rm global}]=0$ with unbounded $S_{\rm gen}[\Sigma_\infty(\mu_Q)]$ at fixed $m$, leading to unbounded violations.

A variation of this idea is to use the generalized entropy in the interior (not the exterior) of the surface $\mu_Q$. It is easy to check that it fails for the same reasons.

\subsection{Possible Modifications of the QPI}
\label{sec-maybe}

We will now discuss an alternative formulation of the QPI that we cannot currently rule out, and we comment on some of its properties that have led us to reject it as our main proposal.

The basic idea is to consider partial Cauchy surfaces other than $L$, still bounded by $\mu_Q$ and remaining inside the black hole. For example, we could assert that
\begin{equation}
m \geq \sqrt{\frac{\hbar S_{\rm gen}[\Sigma]}{4 \pi G}}
\end{equation}\label{eq-QPISigma}
holds for any achronal hypersurface $\Sigma \subset B\cap O_W[\mu_Q]$ whose only boundary is $\mu_Q$. This class includes the lightsheet $L$, so this conjecture would be strictly stronger than our main proposal. It is clear that the heuristic arguments in support of QPI in Sec.~\ref{sec-evidence} also apply to this family of slices.

There are some clear downsides to this choice. The region $B$ and therefore this family of slices are defined teleologically. Furthermore, it is not clear to us how one would formulate a minimality requirement in this case, analogous to the requirement that the classically trapped surface minimize the area on some Cauchy surface.

A variation would be to insist on a Cauchy surface that is as ``long'' as possible, i.e., which does not have any endpoint on the future singularity. Roughly, this means it ends on the future endpoints of the horizon generators, see $\Sigma_{\rm long}$ in Fig.~\ref{fig-slices}. This proposal is weaker than the previous one and neither stronger nor weaker than our main proposal. We will now argue that for an evaporating black hole this results in a less stringent bound than the one obtained from $L$.

As discussed in Sec.~\ref{sec-evidence}, in the Unruh state there is negative entropy falling across the horizon, due to the missing ingoing modes compared to the Hartle-Hawking state. The long slice will capture this negative entropy through the entire process of evaporation. (Here we are assuming that the semiclassical expansion is valid until the black hole area is Planckian in size.) The generalized entropy on this slice is:
\begin{align}
S_{\rm gen}[\Sigma_{\rm long}] = \frac{A[\mu_{Q}]}{4G\hbar} -\gamma \frac{A[\mu_{Q}]}{4G\hbar}~,
\end{align}
where $\gamma\geq 1$ by the GSL, and the second term arises from the contribution of the missing ingoing modes on $\Sigma$.

It is difficult to compute $\gamma$ exactly. If $\gamma > 1$, then $S_{\rm gen}$ will be negative. This renders \eqref{eq-QPISigma} ill-defined. Negative $S_{\rm gen}$ is also conceptually in conflict with the interpretation of $S_{\rm gen}$ as an entropy in the fundamental theory of quantum gravity. This suggests that a careful computation will reveal that $\gamma=1$, in which case Eq.~\eqref{eq-QPISigma} reduces back to the statement of the positivity of the ADM mass. Along with the downsides mentioned earlier, this conundrum shows that such long slices are not ideal for formulating the QPI.

\section{Quantum Penrose Inequality in Anti-de Sitter Space}
\label{sec-ads}

The classical Penrose inequality was motivated by the heuristic argument that a Schwarz\-schild black hole with no exterior matter should have the smallest possible mass for a given trapped surface area. In Eq.~(\ref{eq-pi}) we assumed a vanishing cosmological constant $\Lambda$. An analogous argument for asymptotically Anti-de Sitter spacetimes with curvature scale $L=(-\Lambda/3)^{1/2}$ yields the classical inequality
\begin{equation}
  m \geq f_{\rm AdS}(A[\mu])~,
\label{eq-piads}
\end{equation}
where
\begin{equation}
  f_{\rm AdS}(A)\equiv \left(\frac{A}{16\pi G^2}\right)^{1/2}
  +\left(\frac{A}{16\pi G^2}\right)^{3/2}\frac{G^2}{L^2}~
  \label{eq-fads}
\end{equation}
and $\mu$ is again a trapped surface satisfying an appropriate minimality condition (see Sec.~\ref{sec-CPI}).

\begin{figure}%
\centering
    \includegraphics[width=.475\textwidth]{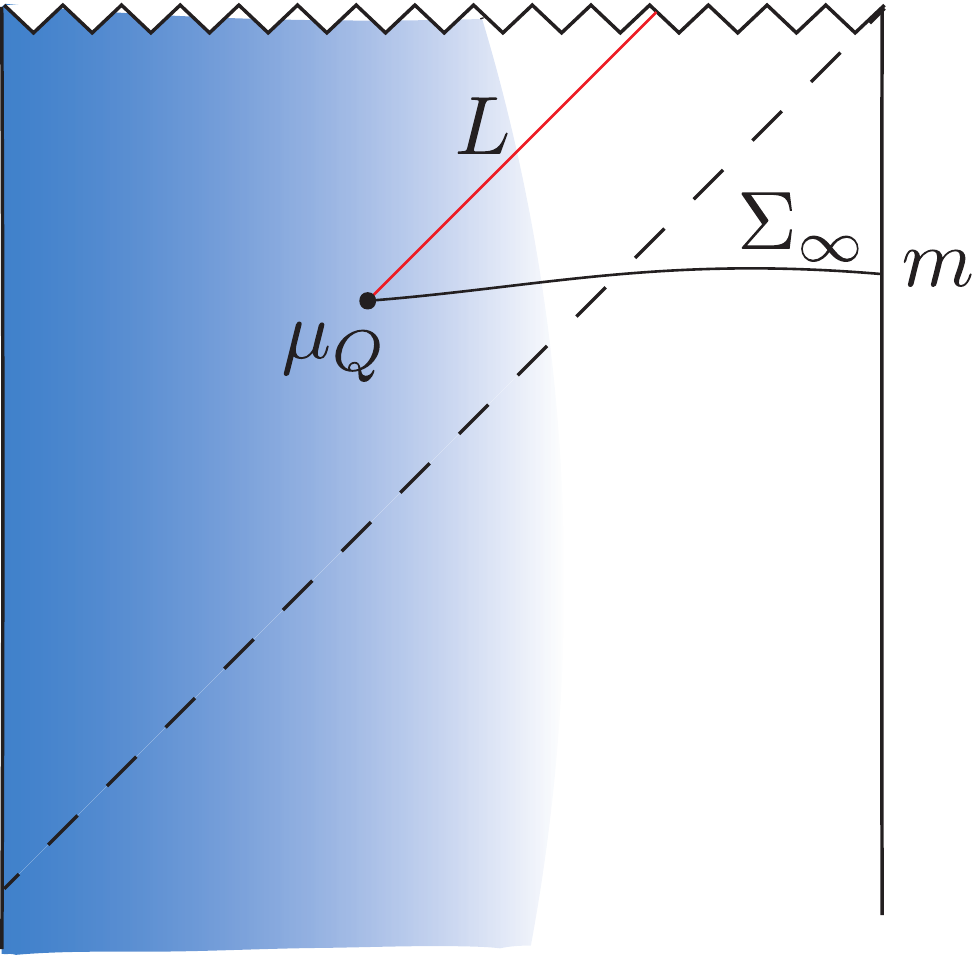}
    \caption{Different choices of slices anchored to the surface $\mu_{Q}$ on which one could compute $S_{\rm gen}$. The red lightsheet $L$ is defined analogously to the asymptotically flat case. Since distant soft modes do not exist for large black holes in AdS, one could also consider computing $S_{\rm gen}$ on the black slice $\Sigma_{\infty}$ that ends on the asymptotic boundary.}%
    \label{fig-AdSslices}%
\end{figure}

\begin{figure}%
    \centering
    \includegraphics[width=.425\textwidth]{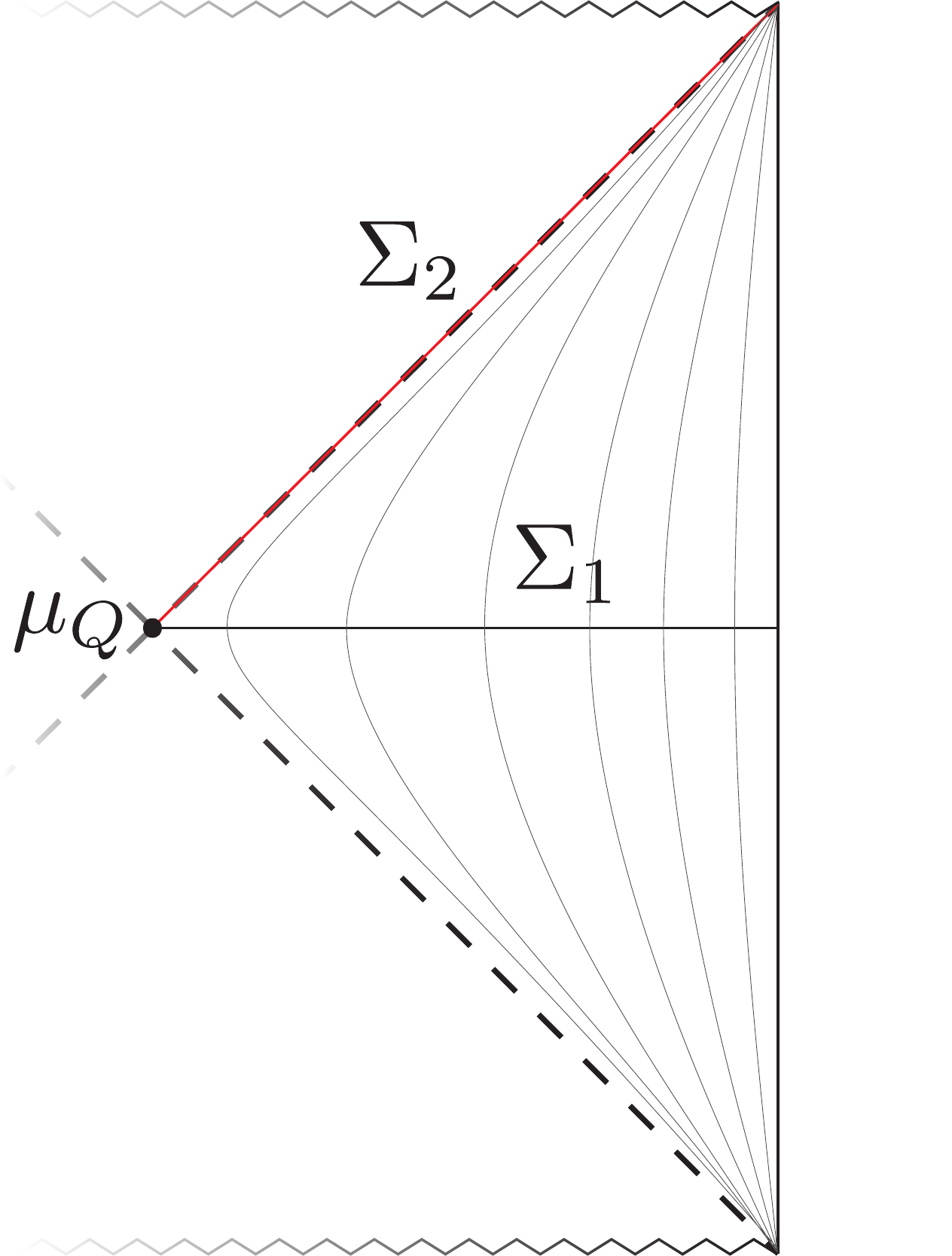}
    \caption{The Hartle-Hawking state is essential for our definition of $f^q_{\rm AdS}$ via $m = f^{q}_{\rm AdS}(S_{\rm gen})$. Here $m$ is the ADM mass including the quantum corrections associated with the radiation mass. $m_{\rm rad}$ is computed on the black slice $\Sigma_{1}$ with respect to the time-like Killing field $t^{\mu}$ whose orbits are shown in the figure. $S_{\rm gen}$ is computed on the red null slice $\Sigma_{2}$ on the horizon that ends on the bifurcation surface $\mu_{Q}$.}%
    \label{fig-adshh}%
\end{figure}

Following our QPI proposal for asymptotically flat space, it would
appear natural to propose the following QPI in asymptotically AdS spacetimes:
\begin{equation}
  m \stackrel{?}{\geq} \left(\frac{\hbar S_{\rm gen}}{4\pi G}\right)^{1/2}
  +\left(\frac{\hbar S_{\rm gen}}{4\pi G}\right)^{3/2}\frac{G^2}{L^2}~.
\label{eq-qpiads}
\end{equation}
in asymptotically AdS spacetimes with curvature scale $L$.  Here $S_{\rm gen}$ is defined with respect to slices defined in Sec.~\ref{sec-formulation}; see Fig.~\ref{fig-AdSslices}.
However, due to $\mathcal{O}(1)$ subtleties discussed at the end of Sec.~\ref{sec-formulation}, it is not clear that Eq.~\eqref{eq-qpiads} will hold exactly in the AdS Hartle-Hawking state (referred to as $\sigma$ henceforth). The issue is the radiation mass outside of the black hole which could be negative, lowering the LHS of Eq.~\eqref{eq-qpiads} to violation. As we will discuss here, in asymptotically AdS spacetimes one could fix this $\mathcal{O}(1)$ issue. Note that the quantum-corrected ADM mass in this state is
\begin{align}
m =  \left(\frac{A}{16\pi G}\right)^{1/2}
  +\left(\frac{A}{16\pi G}\right)^{3/2}\frac{G^2}{L^2} + m_{\rm rad}~,
\label{quantum-mass-AdS}
\end{align}
with
\begin{align}
m_{\rm rad} = \int_{\Sigma_{1}} d\Sigma^{\nu} t^{\mu} \langle T_{\mu\nu} \rangle_{\sigma}~,
\end{align}
where $\langle T_{\mu\nu} \rangle_{\sigma}$ is the renormalized stress tensor in $\sigma$, $\Sigma_{1}$ is a Cauchy slice stretching from the bifurcation surface to the boundary of AdS, and $t^{\mu}$ is the Killing field in Schwarzschild-AdS that is timelike at infinity. Also, note that the area term in Eq.~\eqref{quantum-mass-AdS} is not the quantum-corrected area. Furthermore, based on formulation in Sec.~\ref{sec-formulation}, $S_{\rm gen}$ in the $\sigma$ is computed on the part of the horizon in the future of the bifurcation surface $\mu_{Q}$; see Fig.~\ref{fig-adshh}.
The quantum stress tensor $\langle T_{\mu\nu} \rangle$ has been computed in the Hartle-Hawking state in 2+1 dimensions with different choices of boundary conditions \cite{Steif:1993zv}. One finds that $m_{\rm rad}$ depends on the field content and the boundary conditions; moreover, $m_{\rm rad}$ does not have a definite sign \cite{Steif:1993zv}. Explicit calculations in 2+1 dimensions show that $m_{\rm rad}$ can be negative. We do not expect that the entropy of the matter on $\Sigma_{2}$ and the quantum corrections to the area term would compensate for this negative value of $m_{\rm rad}$ so as to uphold Eq.~\eqref{eq-qpiads}. Therefore, we expect that Eq.~\eqref{eq-qpiads} can be violated in the Hartle-Hawking state. Furthermore, the non-universality of $m_{\rm rad}$ seems to suggest that the correct formulation of QPI for large AdS black holes must depend on various factors that $m_{\rm rad}$ depends on (e.g. the field content and the boundary conditions).

Here we propose a way to introduce this dependence into a Quantum Penrose Inequality for asymptotically Anti-de Sitter spacetimes. Let $f^{q}_{\rm AdS}$ be a function such that in the Hartle-Hawking state,
\begin{align}
  m = f^{q}_{\rm AdS}(S_{\rm gen}[\Sigma_{2}])~,
\label{eq-definef}
\end{align}
where $m$ is the quantum-corrected ADM mass and $S_{\rm gen}$ is associated to the future portion $\Sigma_{2}$ of the horizon; see Fig.~\ref{fig-adshh}. Now, we propose
\begin{align}
  m \geq f^{q}_{\rm AdS}(S_{\rm gen}[L])~,
\label{eq-AdSQPI}
\end{align}
for any marginally trapped surface $\mu_Q$ in an asymptotically AdS spacetime with a large AdS black hole.
A heuristic argument for Eq.~\eqref{eq-AdSQPI} is as follows: First, the above inequality will follow from the classical Penrose inequality unless we are in a state perturbatively close to Kerr-AdS. In that limit, it can be shown (see Appendix~\ref{sec-qscreens}) that given any quantum marginally trapped surface, there exists a Q-screen that approaches the horizon of the Kerr-AdS at late times and has the quantum marginally trapped surface as a leaf. As discussed in Sec.~\ref{sec-QPI}, Q-screens are known to satisfy a generalized second law \cite{BouEng15c}. The QPI would then follow from:
\begin{align}
S_{\rm gen}|_{\rm early} &\leq S_{\rm gen}|_{\rm Kerr-AdS}\nonumber\\
&\implies f^{q}_{\rm AdS} \left(S_{\rm gen}|_{\rm early} \right) \leq f^{q}_{\rm AdS} \left(S_{\rm gen}|_{\rm Kerr-AdS}\right) \leq f^{q}_{\rm AdS} \left(S_{\rm gen}|_{\rm \sigma}\right) = m~,
\label{eq-Sgensketch}
\end{align}
where the first inequality on the second line follows from the generalized second law of Q-screens and the second inequality follows from assuming $f^{q}_{\rm AdS}$ is a monotonic function.

In general we could have states where the AdS black hole is not large enough to reach stable thermal equilibrium with the asymptotic boundary of the spacetime, so a few words about the case of these small AdS black holes are in order. For such black holes, we cannot define the function $f^{q}_{\rm AdS}$ as above. Our proposal would then follow more closely our proposal for asymptotically flat spacetimes, where we formulate our conjecture using the function $f$ appearing in the classical Penrose inequality for AdS. 
\begin{align}
m \geq f_{\rm AdS}(4G\hbar S_{\rm gen}[L])~,
\label{QPIAdSsmall}
\end{align}
where $f_{\rm AdS}$ is defined in Eq.~\eqref{eq-fads}.
The phase transition for (in)stability of AdS black holes happen around ADM mass $L/G$, so our proposal changes for mass above and below the phase transition point. The exact value of mass associated to a phase transition depends on the choice of boundary conditions and the field content.\\

An important difference between QPI for large AdS black holes and flat space black holes is the absence of the challenge associated with soft modes. As discussed in Sec.~\ref{sec-alternatives}, in asymptotically flat space, one can add entropy far away from the black hole at negligible cost to the ADM mass. This prevents any formulation of QPI where the generalized entropy is computed on partial Cauchy slices approaching spatial infinity in asymptotically flat spacetimes.

However, in asymptotically AdS spacetimes and in the presence of a large black hole, excitations require considerable energy  to remain outside of the black hole, so the arguments of Sec.~\ref{sec-alternatives} do not go through and matter entropy outside of the black hole has an energy cost. Therefore, in the presence of large AdS black holes the slice on which $S_{\rm gen}$ is evaluated could end on the asymptotic boundary of AdS. This possibility was discussed in the context of AdS/CFT in~\cite{EngHor19}. To define the function $f^{q}_{\rm AdS}$ in this version of QPI, we need to consider the Hartle-Hawking state and the generalized entropy on the spatial slice $\Sigma_{1}$ of Fig.~\ref{fig-adshh},
\begin{align}
  m = f^{q}_{\rm AdS}(S_{\rm gen}[\Sigma_{1}])~.
\label{eq-definef-new}
\end{align}
The quantum extremal surface prescription~\cite{EW14} equates $S_{\rm gen}[\Sigma_{1}]$ with the von Neumann entropy of the dual CFT in the thermofield double state. Therefore, this definition of the function $f^{q}_{\rm AdS}$ has a very natural interpretation from the CFT perspective
\begin{align}
\langle H \rangle_{TFD} = f^{q}_{\rm AdS}\left(S_{\rm CFT}[TFD]\right)~,
\end{align}
where $\langle H \rangle_{TFD}$ is the expectation value of the CFT Hamiltonian in the thermofield double state.

\section{Classical and Non-gravitational Limits}
\label{sec-limits}

In this section we discuss two interesting limits of the QPI: the classical limit, $\hbar \to 0$; and the non-gravitational limit, $G \to 0$.

In the $\hbar \to 0$ of QPI, we recover the classical Penrose inequality. This is easy to see. The amount of matter entropy on $L$ is $\mathcal{O}((G\hbar)^{0})$, and therefore
\begin{align}
\lim_{\hbar \to 0}4G\hbar S_{gen}[L] = A[\mu_{Q}]~.
\end{align}
Furthermore, the surface $\mu_{Q}$ is perturbatively close to a (classically) marginally trapped surface such that their area difference is due to quantum corrections and therefore of order $G\hbar$ and can be neglected. Lastly, any $\hbar$ corrections to the function $f$ can trivially be ignored in the $\hbar \to 0$ limit. We therefore have the desired implication:
\begin{align}
f^{q}\left(4G\hbar S_{gen}[\mu_{Q}]\right) \leq m \stackrel{\hbar \to 0}{\implies} f^{c}\left(A[\mu]\right) \leq m~.
\label{eq-QPItoPI}
\end{align}

We turn to the $G\to 0$ limit of the QPI. This is of interest because some semiclassical conjectures yield nontrivial and novel implications about QFT in this limit. For example, the QNEC was first discovered by taking the $G\to 0$ limit of the QFC in a particular setting~\cite{BouFis15a}. In order to sidestep the small ``Casimir uncertainty'' discussed in Sec.~\ref{sec-formulation}, we will consider the QPI in AdS. We further restrict to two complementary scenarios.

First, consider a perturbation to a Hartle-Hawking state such that in a finite amount of time the state settles back down to a Hartle-Hawking state (with a different temperature). In this case, Eq.~\eqref{eq-Sgensketch} shows that the QPI is equivalent to the GSL. The non-gravitational limit of the GSL the monotonicity of relative entropy. This is a nontrivial but well-known statement in quantum information theory, which applies in particular in QFT.

The second scenario is when the perturbation does not relax to equilibrium. This means that the excitation that takes the state away from the Hartle-Hawking state remains outside of the black hole. Therefore such excitations do not change the generalized entropy on $L$  or the geometry of the event horizon. Let $\delta m$ be the change in the ADM mass caused by this perturbation. Since the QPI is saturated in the Hartle-Hawking state, it reduces to
\begin{align}
\delta m \geq 0
\label{eq-Gtozero2}
\end{align}
in the $G\to 0$ limit. Here $\delta m = \int_{\Sigma_{1}} d\Sigma^{\mu} \xi^{\nu} T_{\mu\nu}$ (see Fig.~\ref{fig-adshh}), and $t^{\nu}$ is the timelike Killing vector field outside the black hole. This makes physical sense: if the field excitations are isolated from the black hole, they need to satisfy their own positive energy condition.

\section{Cosmic Censorship Conjecture}
\label{sec-ccc}

In this section, we consider the current status of the cosmic censorship conjecture (CCC) and its relation to the Penrose inequality. We argue that there is a need for a quantum generalization of the CCC, and we suggest that the proposed Quantum Penrose Inequality may inform the formulation of a quantum CCC.

The formation of singularities in gravitational collapse is guaranteed by classical~\cite{Pen64} and quantum~\cite{Wal13} singularity theorems. However, it is not clear that the formation of a singularity implies the formation of a black hole.

The weak CCC asserts that singularities (regions of arbitrarily high curvature) will not be visible to a distant observer.\footnote{We will consider only the weak CCC here. The strong form of the CCC states, roughly, that no observer can see a singularity. In all cases, one assumes regular initial data.}
A precise statement of the conjecture can be formulated as follows~\cite{Wald}:
\textit{Let ($\Sigma,$ $h_{\mu\nu},$ $K_{\mu\nu} $) be an asymptotically flat initial data set for Einstein's equation with ($\Sigma$, $h_{\mu\nu}$) a complete Riemannian manifold. Let the matter sources be such that $T_{\mu\nu}$ satisfies the dominant energy condition and the coupled Einstein-matter field equations are of the form $\displaystyle \Box \phi(x) = F(x, \phi, \nabla_{\mu}\phi) $, where $F$ is a smooth function of its variables. In addition let the initial data for the matter fields on $\Sigma$ satisfy appropriate asymptotic falloff conditions at spatial infinity. Then the maximal Cauchy evolution of these initial data is an asymptotically flat, strongly asymptotically predictable spacetime.}

The CCC has not been proven. Indeed, there are a number of known ``mild'' violations that we will discuss shortly. The (classical) Penrose inequality is only a necessary condition for the CCC, as explained in Sec.~\ref{sec-CPI}. Even this weaker statement has not been proven; but as a quantitative relation between mass and area, it has been extensively explored. The fact that no counterexample has been found can be viewed as indirect evidence that some version of the CCC may indeed hold.

Let us now discuss the mild violations mentioned in the previous paragraph. A black string in 4+1 dimensions suffers from the Gregory-Laflamme instability \cite{GregoryLaflamme1993}. Further evolution causes the string to become arbitrarily thin in some regions~\cite{2011arXiv1107.5821G, 2011arXiv1106.5184L} and so arbitrarly high curvatures become visible to a distant observer.

In 3+1 dimensions, there exist fine-tuned initial data sets such that the solution exhibits a self-similar behavior near the threshold of formation of a black hole~\cite{WaldCCC1997, Gundlach2007, Choptuik1993, Christodoulou1984}. At the threshold, a naked singularity forms. In some examples, the naked singularity propagates out to ${\cal I}^+$.

In the above two examples, the initial data satisfy the dominant energy condition, as required by the CCC. The black string is not asymptotically flat, but one expects that it can be truncated at a sufficiently great length so that local evolution far from the ends still leads to a naked singularity.

Let us now add a third example, which is physically relevant but does not obey the dominant energy condition: a black hole that evaporates completely. In this case, treating the spacetime as a classical manifold, a naked singularity is inevitable~\citep{1979PThPh..62.1434K, Wald84a}.

Only the last example explicitly involves quantum effects. But it points to a resolution of all three violations: clearly, it makes no sense to treat the spacetime as a classical manifold near the endpoint of evaporation (i.e., arbitrarily close to the naked singularity). When the curvature formally exceeds the Planck curvature, the semiclassical expansion breaks down, and a classical geometric description of the spacetime need not exist.

But this observation also applies to the other known examples of CCC violation. One would expect a black string to pinch off before it becomes thinner than a Planck length. Similarly, one would expect quantum effects to smooth out the fine-tuned initial data, or at least the singularities they lead to.

Naively, all three examples violate the spirit of the CCC: starting from a highly classical regime, evolution produces an outcome in which quantum gravity is required to maintain predictability. But in an important sense, the violation is ``small'' in each case. The energies involved are likely no greater than the Planck mass, and we can ``guess'' a plausible future evolution without having a full quantum gravity theory. For example, the Planck-sized black hole will probably decay into a few more particles, and the black string will simply pinch off.

It would be of interest to formulate a quantum version of the CCC which accounts for these physically reasonable phenomena, i.e., one that is not formally violated by them.\footnote{A specific proposal will be studied in forthcoming work.} We expect the Quantum Penrose Inequality to play a role analogous to the classical one: as a necessary condition for the quantum CCC, and thus as a useful test. Perhaps more importantly, the Quantum Penrose Inequality may be of some use in identifying the correct formulation of a quantum CCC in the first place.

\acknowledgments

We thank Roberto Emparan, Netta Engelhardt, Gary Horowitz, Donald Marolf,  Robert Myers, and Aron Wall for discussions. This work was supported in part by the Berkeley Center for Theoretical Physics; by the Department of Energy, Office of Science, Office of High Energy Physics under QuantISED Award DE-SC0019380 and contract DE-AC02-05CH11231; and by the National Science Foundation under grant PHY1820912. MT acknowledges financial support coming from the innovation program under ERC Advanced Grant GravBHs-692951. 

\appendix

\section{(Quantum) Trapped Surfaces in the Schwarzschild Geometry}
\label{sec-geom}

\subsection{Classical Solution and Semiclassical Corrections}
\label{sec-classical}

The Schwarzschild metric is
\begin{equation}
  ds^2 = -\left(1-\frac{R}{r}\right) dt^2 + \frac{dr^2}{1-R/r} + r^2 d\Omega^2~.
\end{equation}
where $R=2GM$ is the Schwarzschild radius. In ingoing Eddington-Finkelstein coordinates,
\begin{equation}
  ds^2 =  -\left(1-\frac{R}{r}\right) dv^2 + 2dv\,dr + r^2 d\Omega^2~,
\end{equation}
where
\begin{equation}
  v=t+r_*~,~~ r_* = r+R \log|\frac{r}{R}-1|~,~~ \frac{dr}{dr_*}=1-\frac{R}{r}~.
\label{eq-tort}
\end{equation}

Ingoing radial null congruences are at constant $v$, so $dv=0$. Outgoing null congruences satisfy $dv = 2 dr_*$, so
\begin{equation}
  v= 2 r_* + \text{ const}~.
\label{eq-outv}
\end{equation}
We are interested in their expansion,
\begin{equation}
  \theta=\frac{dA/d\lambda}{A}
\end{equation}
in terms of a convenient affine parameter, $\lambda$.

To find $\lambda$, first note that $r$ is an affine parameter. This follows because $A=4\pi r^2$, so
\begin{equation}
  \theta = \frac{2}{r} \frac{dr}{d\lambda}~;
  \label{eq-thetar}
\end{equation}
and Raychaudhuri's equation in the vacuum, for spherical symmetry, reduces to
\begin{equation}
\frac{d\theta}{d\lambda}+\frac{1}{2}\theta^2=0~.
\end{equation}
This implies that $dr/d\lambda$ must be constant for any affine $\lambda$. We can take that constant to be 1 if we like, and choose another constant of integration so that $r=\lambda$.

However, this choice is not convenient for outgoing lightrays, because we are interested in radial null congruences near and on the event horizon,
\begin{equation}
  |r-R|\ll R~.
\label{eq-near}
\end{equation}
Intuitively, the radius $r$ does not change much for these congruences, so small changes in $r$ correspond to large motions along the congruence. On the horizon, $r$ is degenerate, and inside the black hole, $r$ runs towards the past.

To remedy this, let us consider the coordinate distance $c=r-R$ from the horizon. We will work in the near-horizon limit of Eq.~(\ref{eq-near}), i.e., to first order in $c/R\ll 1$.  For example, $r_* = R+R\log (|c|/R)$ in this approximation; and by Eq.~(\ref{eq-outv}), an outgoing congruence satisfies $v=2R \log (|c|/R)+$ const.
Inverting this, we find
\begin{equation}
  c=c_0 e^{v/2R}
\end{equation}
where $c_0$ is the coordinate distance from the horizon at $v=0$. This is the quantity that vanishes on the horizon and goes negative inside, so we can define a nondegenerate, always future-directed parameter by choosing $\lambda= c/c_0$. This is affine since $\lambda=(r-R)/c_0$ and $r$ is affine.

To summarize, we choose the affine parameter
\begin{equation}
  \lambda = e^{v/2R}
  \label{eq-lambdav}
\end{equation}
on outgoing null geodesics near the horizon. By Eq.~(\ref{eq-thetar}), the expansion of any such congruence is given by
\begin{equation}
  \theta =\frac{2c_0}{R}~,
  \label{eq-tc0}
\end{equation}
where we again used $r-R\ll R$. All surfaces on the event horizon have
$c_0=0$ and hence $\theta=0$; they are marginally outer trapped. It is easy to check that these are the only such surfaces.

Any null vector tangent to the outgoing congruences must be proportional to $\partial_t+\partial_{r_*}$. Let $k^a$ be the particular null vector associated to the affine parameter $\lambda$. From Eq.~(\ref{eq-lambdav}) we have 
\begin{equation}
  k=\frac{d}{d\lambda} = \frac{2R}{\lambda} \left. \frac{d}{dv}\right|_{\text{cong}} = \frac{R}{\lambda} (\partial_t + \partial_{r_*})~,
\end{equation}
For the second equality, we used that on the outgoing congruence $t = (v+ \text{const})/2$, $r_* = (v- \text{const})/2$.

For all ingoing spherical congruences in the region covered by the
ingoing Eddington-Finkelstein coordinates, $-r$ is a future-directed
nondegenerate affine parameter. Thus Eq.~(\ref{eq-thetar}) implies that their expansion, $\theta_l$, is everywhere negative. This establishes that every spherical cut of the event horizon is marginally trapped, i.e., satisfies $\theta=0$ and $\theta_l\leq 0$.


To treat quantum matter as a small perturbation, we expand the Einstein equation, $G_{ab} = 8\pi G \langle T_{ab} \rangle$, in powers of $G\hbar$, to first order. (We drop the expectation value symbol below.) In this approximation, we can compute matter effects on the expansion of congruences by integrating the Raychaudhuri equation,
\begin{equation}
  \frac{d\theta}{d\lambda} = -\frac{1}{2}\theta^2 - \varsigma^2 - 8\pi G T_{kk}~.
\label{eq-raych}
\end{equation}
Here $T_{kk} = T_{ab} k^a k^b$, and $k^a = (\frac{d}{d\lambda})^a$ is the affine tangent vector to the null congruence. The shear term vanishes for the spherical congruences we consider. In general, the $\theta^2$ term will be $O((G\hbar)^0)$ and thus dominant.

However, here we will be interested in surfaces where classical and quantum effects compete. Such surfaces must have $\theta\sim O(G\hbar)$ classically. By Eq.~(\ref{eq-tc0}) they are found in a neighborhood $|c|\leq O(G\hbar)$ of the event horizon. Hence $\theta^2\sim O((G\hbar)^2)$ will be negligible in the region of interest, and Eq.~(\ref{eq-raych}) reduces to
\begin{equation}
  \theta(\lambda)-\theta(\lambda_0) = -8\pi G\int_{\lambda_0}^\lambda T_{kk}~.
  \label{eq-int}
\end{equation}

\subsection{Classically Trapped Surfaces During Evaporation}
\label{sec-cts}

We will now compute the effect of the quantum stress tensor for the Unruh state \cite{Candelas} on the position of (marginally) trapped surfaces in the Schwarzschild geometry.

The renormalized stress tensor in the Unruh vacuum takes the form
\begin{equation}
    \bra{U}T_{a}^{\; b}\ket{U}_{\text{ren}} \xrightarrow{r \xrightarrow{} 2M} \frac{L}{4\pi R^2}  \begin{pmatrix}
    f^{-1} & -1 \\
    f^{-2} & -f^{-1} 
  \end{pmatrix},
\end{equation}
where $\displaystyle f = (1 - R/r)$, $R = 2M$, $a$ and $b$ range over $t$ and $r$, and
\begin{equation}
  L\sim \frac{\hbar}{R^2}
\end{equation}
is the luminosity of the black hole. Lowering indices we find 
\begin{equation}
    \bra{U}T_{ab}\ket{U}_{\text{ren}} \xrightarrow{r \xrightarrow{} 2M} \frac{L}{4\pi R^2}  \begin{pmatrix}
    -1 & -f^{-1} \\
    -f^{-1} & -f^{-2} 
  \end{pmatrix},
\end{equation}
Using
\begin{equation}
    \partial_{r*} = \frac{dr}{dr*}\partial_r = \bigg(1 - \frac{R}{r}\bigg)\partial_r~,
\end{equation}
we can express the null vector $k$ in $(t,r)$ coordinates,
\begin{equation}
    k = \frac{R}{\lambda}\bigg(\partial_t + \bigg(1 - \frac{R}{r}\bigg)\partial_r\bigg) = k^t\partial_t + k^r\partial_r~.
\end{equation}
and we obtain
\begin{equation}
    \begin{split}
        \expval{T_{\mu\nu}k^{\mu}k^{\nu}} & = \expval{T_{tt}k^tk^t} + \expval{T_{rr}k^rk^r} + 2\expval{T_{tr}k^tk^r}  \\
        & = - \frac{L}{\pi \lambda^2} = -\frac{\hbar}{\pi R^2\lambda^2}
    \end{split}
\end{equation}

Next we compute the change in the expansion induced by the above quantum stress tensor. We consider a black hole at the onset of evaporation, for which there is no Hawking radiation outside the near horizon zone yet. Thus we expect the geometry to revert to the classical vacuum Schwarzschild solution far from the black hole. And so, to find the corrected expansion, we integrate backwards from $\lambda=\infty$ to find the shift:
\begin{equation}
\begin{split}
    \delta \theta \equiv \theta(\lambda) - \theta(\infty) & = -8\pi G\int_{\infty}^{\lambda} \expval{T_{\mu\nu}k^{\mu}k^{\nu}} d\lambda' = \\
    & = 8\pi G\int_{\lambda_0}^{\lambda} \frac{\hbar}{\pi R^2\lambda'^2}d\lambda' = - \frac{8 G\hbar}{ R^2\lambda}~.
\end{split}
\end{equation}

To find the (classically) marginally trapped surfaces in the Unruh state, we solve
\begin{equation}
  \theta^{(0)}+\delta\theta=0~,
\end{equation}
where $\theta^{(0)}$ is the uncorrected classical expansion given in Eq.~\ref{eq-tc0}. Using $c=c_0\lambda$, we find that the classical marginally trapped surfaces are located at
\begin{equation}
  c_\text{MTS}\sim \frac{G\hbar}{R}
  \label{eq-qct1}
\end{equation}
in the quantum-corrected geometry. 
Very near the horizon, we can treat the radial coordinate to be essentially $R$ to zeroth order. 

An alternative useful notion of distance is the proper radial distance from the horizon, $\ell$, which satisfies
\begin{equation}
    d\ell = \frac{dr}{\sqrt{1 - \frac{R}{r}}} \simeq \sqrt{R}\frac{dr}{\sqrt{r - R}}\hspace{10pt}\xrightarrow{}\hspace{10pt} \ell \simeq 2\sqrt{R(r - R)} \sim (Rc)^{1/2}
\end{equation}
Since $G\hbar=l_p^2$, we see that the trapped surfaces are about a Planck length outside the horizon:
\begin{equation}
    \ell_\text{MTS} \sim \order{l_p}.
\end{equation}
Thus, the area of the classical marginally trapped surface is increased by the quantum correction, by
\begin{equation}
  \Delta A_\text{MTS} \sim G\hbar =l_P^2
\end{equation}

\subsection{Quantum Trapped Surfaces During Evaporation}
\label{sec-qts}

We still consider the quantum-corrected geometry in the Unruh state, so the classical expansion is given by
\begin{equation}
  \theta = \theta^{(0)}+\delta\theta \sim \frac{c_0}{R}- \frac{G\hbar}{R^2\lambda}~.
  \label{eq-totaltheta}
\end{equation}
The generalized entropy is
\begin{equation} 
  S_\text{gen} = \frac{A}{4G\hbar}+S~,
\end{equation}
where $S=- \text{Tr}\, \rho\log\rho$ and $\rho$ is the quantum state in the region exterior to the Cauchy-splitting sphere. The quantum expansion $\Theta$ is ($4G\hbar$ times) the rate of change of the generalized entropy, per unit area, under shape deformations. In the spherically symmetric case,
\begin{equation}
  \Theta = \theta + \frac{4G\hbar}{A}\frac{dS}{d\lambda}~,
  \label{eq-bigth}
\end{equation}
Quantum marginally trapped surfaces are characterized by $\Theta=0$. 

The Generalized Second Law (GSL) states that any outgoing radial congruence on or outside the event horizon must satisfy $\Theta\geq 0$, so the quantum marginally trapped surfaces must lie inside the horizon \cite{Wal13}. By Eq.~(\ref{eq-totaltheta}), $\theta<0$ on and inside the horizon.   We see from Eq.~(\ref{eq-bigth}) that the GSL requires
\begin{equation}
  \frac{4G\hbar}{A} \frac{dS}{d\lambda} = -\alpha \theta|_{\mathcal{H}}~,
  \label{eq-dsda}
\end{equation}
where $\mathcal{H}$ refers to the horizon. We take $\alpha-1\sim O(1)$, in line with Page's explicit calculation for an evaporating black hole in the Unruh state~\cite{Pag76}.

Combining these results and neglecting factors of order unity where appropriate, we find
\begin{equation}
  \Theta = \theta - \alpha\theta|_{\mathcal{H}} 
     = \frac{c}{R\lambda} - \frac{G\hbar}{R^2\lambda} + \alpha\frac{G\hbar}{R^2\lambda}~.
\end{equation}
Setting $\Theta=0$ yields
\begin{equation}
    \frac{c}{R\lambda} = -(\alpha - 1)\frac{G\hbar}{R^2\lambda} \hspace{10pt}\xrightarrow{}\hspace{10pt} c \sim -\frac{G\hbar}{R}.
\end{equation}
Using the proper area, we find
\begin{equation}
  \Delta A_\text{QMTS} \sim -l_P^2~.
\end{equation}
Thus, the quantum marginally trapped surfaces are a proper distance of order the Planck length inside of the horizon.

\begin{figure}%
    \centering
    \includegraphics[width=.7\textwidth]{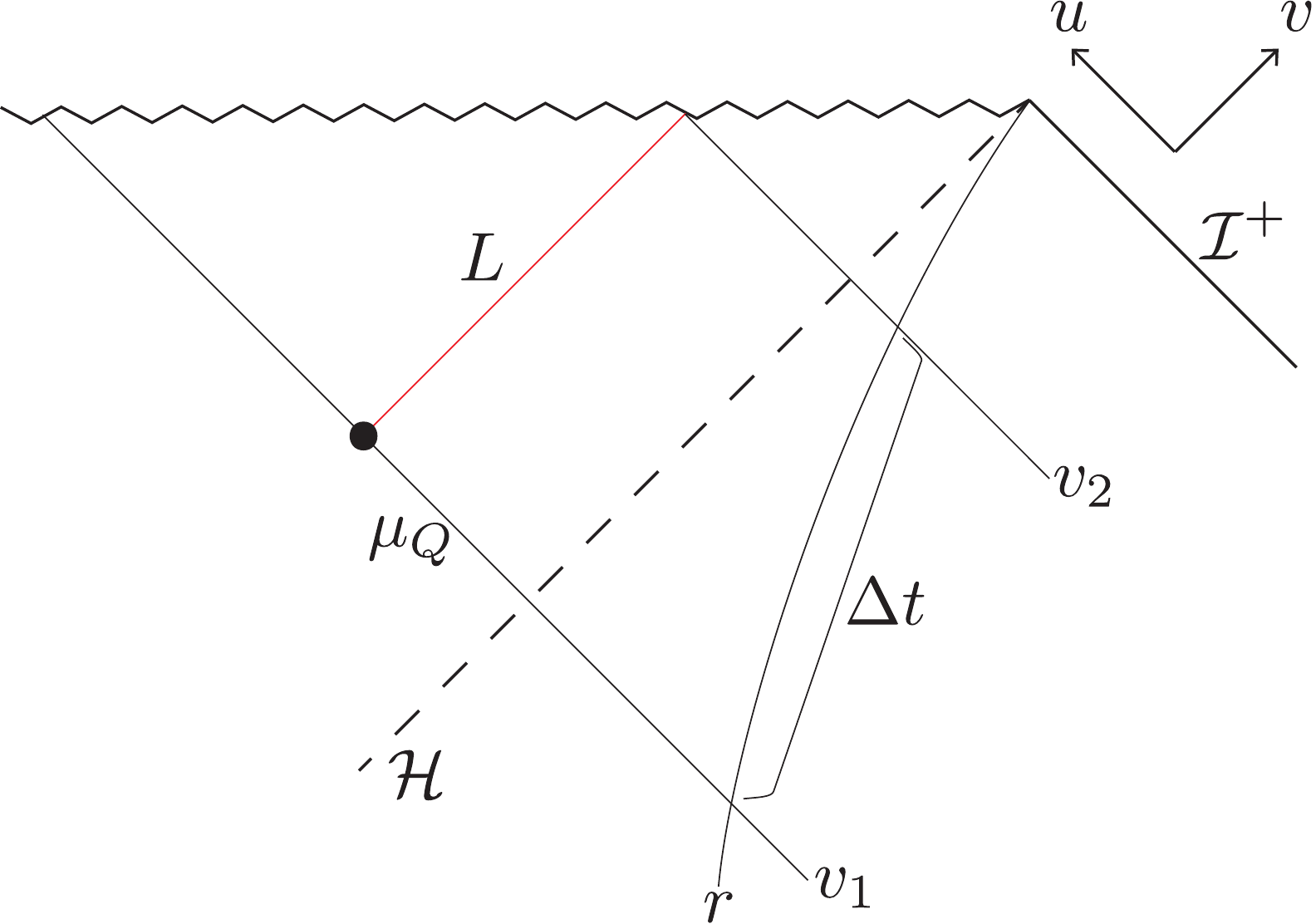}
    \caption{The future outgoing lightsheet of $\mu_Q$ (top red line) is crossed by two ingoing radial null geodesics at $v_1$ (at $\mu_Q$) and $v_2$ (at the singularity). Their Schwarzschild time difference at fixed $r$ is the scrambling time, $\Delta t_s$.}%
    \label{fig-scramblingtime}%
\end{figure}
%

We will now show that the ``duration'' of the lightsheet $L$ of a quantum marginally trapped surface $\mu_Q$ is of order of scrambling time \begin{equation}
  \Delta t_s\sim R\log\frac{R}{l_P}~.
\end{equation}
This assumes that $\mu_Q$ is about one Planck length inside of the event horizon, as would be the case for an isolated, slowly evaporating black hole. Of course, the points on $L$ are null or spacelike separated. What we mean by the ``duration'' of $L$ is the amount of time, as measured at large radius $r$, for which it will be the case that matter falling in radially from this radius will cross $L$ (see Fig.~\ref{fig-scramblingtime}).

We will approximate the infalling matter as ingoing radial null geodesics; the result would be the same for timelike geodesics starting at rest at large radius. Let the earliest geodesic crossing $L$ be at $v=v_1$ in the Eddington-Finkelstein coordinates defined in Appendix~\ref{sec-classical}. It will meet $L$ at $\mu_Q$, whose radius satisfies $R-r_{\mu_Q}\sim l_P^2/R$. The last geodesic that meets $L$ will do so where $L$ hits the singularity, at $r=0$. The lightsheet $L$ is characterized by $u = const$, where $u$ is the ingoing Eddington-Finkelstein coordinate, $u\equiv t - r_*$. Here $r_*$ is the tortoise coordinate defined in Eq.~\ref{eq-tort}. Since $r_*$ depends only on $r$, we have
\begin{equation}
  \Delta t = t_2-t_1=r_*(r_{\mu_Q})-r_*(0) = r_{\mu_Q}+R\log\frac{R}{l_P^2/R} \sim \Delta t_s~.
\end{equation}

A similar analysis demonstrates that the scrambling time is how long it takes a geodesic to propagate from about a Planck distance outside the horizon to the edge of the near-horizon zone, at $r=3R/2$. 

\section{Perturbative Construction of Q-screens}
\label{sec-qscreens}

{\em Let $\mu_Q$ be a quantum marginally trapped surface near a perturbed Killing horizon that approaches the Hartle-Hawking state in the future. Then there exists a Q-screen that approaches the Killing horizon in the future and contains $\mu_Q$ as a leaf.}

This fact is useful in sketching a heuristic argument for our conjectured QPI in asymptotically AdS spacetime, following Eq.~\eqref{eq-Sgensketch}. We will now demonstrate this claim by explicit construction. 
\begin{figure}%
    \centering
    \includegraphics[width=.425\textwidth]{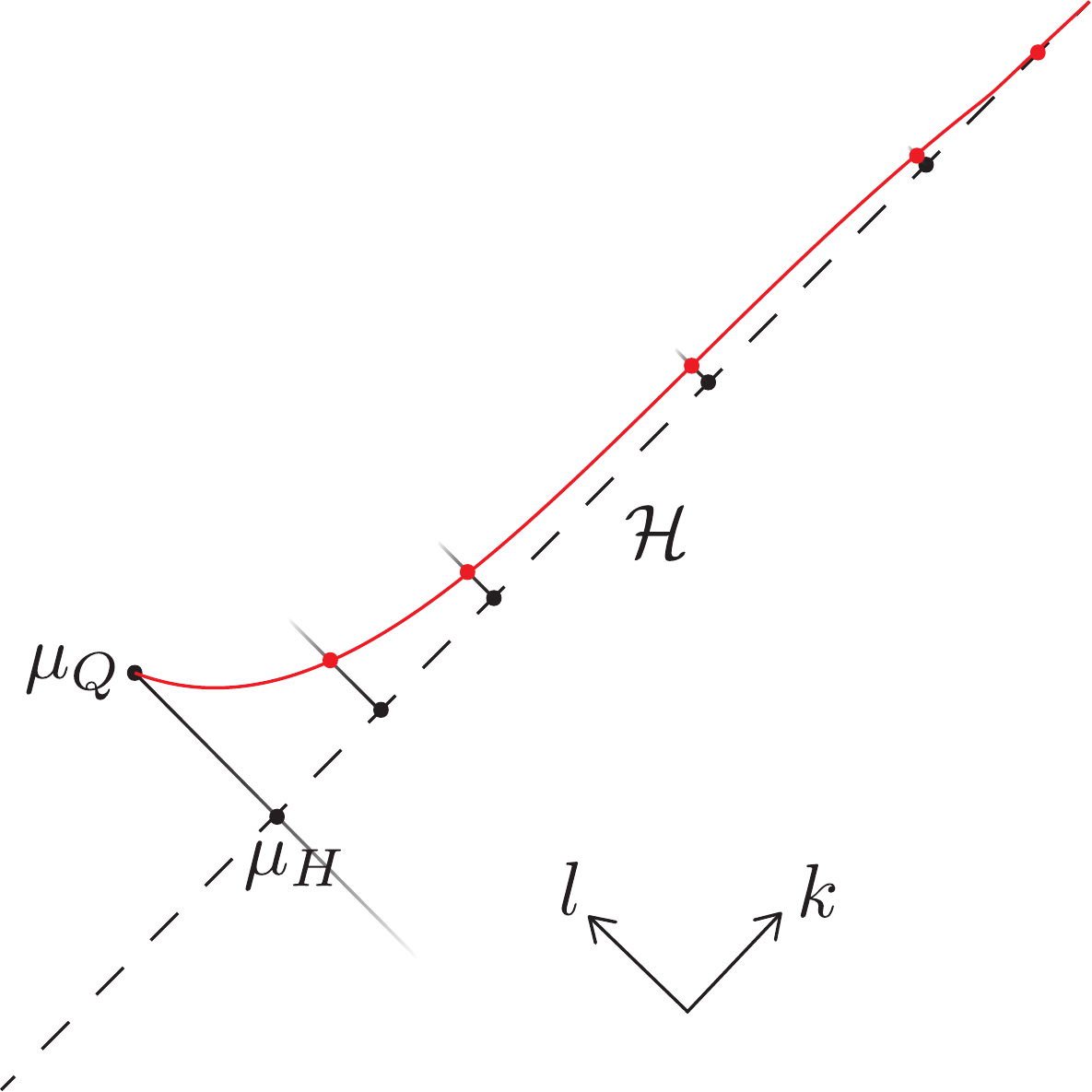}
    \caption{A quantum marginally trapped surface $\mu_{Q}$ in the vicinity of a perturbed Killing horizon $\mathcal{H}$. We construct a Q-screen containing $\mu_{Q}$ that asymptotes to the Killing horizon at late times. We first fire a null plane towards $\mathcal{H}$ that intersects it on $\mu_{H}$. We then foliate $\mathcal{H}$ starting from $\mu_{H}$. At every leaf of this foliation, we fire null planes inwards and to the future. On each null plane, we find a quantum marginally trapped surface at an affine distance $\delta U$ from $\mathcal{H}$. The Q-screen is the union of these quantum marginally trapped surfaces.}%
    \label{fig-Qscreen}%
\end{figure}

Consider an event horizon $\mathcal{H}$ which is a perturbation to a Killing horizon caused by matter excitations $T_{\mu\nu} \sim \mathcal{O}(\hbar)$ such that in the far future $\mathcal{H}$ settles down to a Killing horizon in the Hartle-Hawking state. Furthermore, assume that there exists a quantum marginally trapped surface near $\mathcal{H}$. It is known \cite{Wal13} that quantum marginally trapped surfaces are behind event horizons, so $\mu_{Q}$ will be a small distance in the inward direction $l$ from $\mathcal{H}$. Given any co-dimension 2 surface in this spacetime, $k$ and $l$ respectively represent the outward and inward null vectors perpendicular to the surface. Let $y$ parametrize the transverse position of the surface; see Fig. \ref{fig-Qscreen}.

For the construction of the Q-screen, we start by emanating a past outwards-directed null plane from $\mu_{Q}$ and mark its intersection with the horizon as $\mu_{H}$. Now, we can pick a foliation of the horizon that starts from $\mu_{H}$ and continues towards the future of $\mathcal{H}$ such that it eventually approaches the preferred foliation of the Killing horizon. Mark the leaves of this foliation by $\lambda$ such that $\lambda = 0$ is $\mu_{H}$ and $\lambda$ grows along the future leaves. We construct the Q-screen by shooting null future-directed inward null planes from the leaves $\mu_{H}$ and on that null plane look for a quantum marginally trapped surface.

Suppose that a given leaf of our foliation of $\mathcal{H}$ (marked by $\lambda$) has a quantum expansion $\Theta_{k}(\lambda;y)$ at a given transverse position $y$. By the generalized second law, $\Theta_{k}\geq 0$. Then, perturbatively we can find the location of a quantum marginally trapped surface as
\begin{align}
\Theta_{k} (\lambda;y) + \delta U(\lambda; y) \left(\partial_{l} \Theta_{k}(\lambda;y)\right) = 0~,
\label{eq-pertThetak}
\end{align}
where $\delta U$ is the amount of affine parameter in the $l$ direction we need to venture to find a quantum marginally trapped surface and $\Theta_{k} = \mathcal{O}(G\hbar)$.

We need to solve for a function $\delta U(y)$ and show that it approaches zero as we go towards higher values of $\lambda$. From the definition of quantum expansion it follows that
\begin{align}
\partial_{l} \Theta_{k} = \partial_{l} \theta_{k} + 4G\hbar \partial_{l} \partial_{k} S_{\rm out}~.
\end{align}
The cross-focusing equation is
\begin{align}
\partial_{l} \theta_{k} = - \frac{1}{2} \mathcal{R} - \theta_{l} \theta_{k}+ \nabla . \chi+ \chi^{2} + 8\pi G~T_{kl}~,
\end{align}
where $\mathcal{R}$ is the intrinsic Ricci scalar of the leaf and $\chi$ is its twist~\cite{BouMoo16}. From Eq.~\eqref{eq-pertThetak}, we see that in order to solve for $\delta U$ to first non-trivial order in $G\hbar$, we only need the leading order expression for $\partial_{l} \Theta_{k}$. The leading order term is
\begin{align}
\partial_{l} \Theta_{k} = - \frac{1}{2} \mathcal{R}^{(0)} + \mathcal{O}(G\hbar)~,
\end{align}
where $\mathcal{R}^{(0)}$ is the ($y$-independent) intrinsic Ricci scalar of the leaf on the unperturbed Killing horizon. For a 2-sphere $\mathcal{R}^{(0)} = 2$. Combining the above equations with \eqref{eq-pertThetak}, we can solve for $\delta U$ to the first non-trivial order in $\mathcal{O}(G \hbar)$:
\begin{align}
\delta U(y;\lambda) = \Theta_{k} (\lambda;y) ~.
\label{eq-deltaU}
\end{align}
Since by assumption $\mathcal{H}$ approaches a Killing horizon in the Hartle-Hawking state in the future, we have
\begin{align}
\lim_{\lambda \to \infty} \Theta_{k} (\lambda;y) = 0 \implies \lim_{\lambda \to \infty} \delta U(\lambda;y) =0~,
\end{align}
where the implication follows from Eq.~\eqref{eq-deltaU}. This means that the leaves of the Q-screen start at $\mu_{Q}$ and approach the late times of the event horizon, which is what we set out to show.

\bibliographystyle{utcaps}
\bibliography{all}

\end{document}